\documentclass[sigconf, nonacm]{acmart}
\AtBeginDocument{%
  }

\usepackage{multirow}
\usepackage{multicol}
\usepackage{enumitem} 
\usepackage{booktabs}
\usepackage{listings}
\usepackage{xcolor}
\usepackage{color}
\usepackage{soul}
\usepackage{cleveref}
\usepackage{tabularx}
\usepackage{xtab}
\usepackage[table]{xcolor} 
\usepackage{longtable}
\usepackage[most]{tcolorbox}
\usepackage[colorinlistoftodos, textsize=small]{todonotes}
\setuptodonotes{inline}

\definecolor{cellc1}{HTML}{F0F4F8}
\definecolor{cellc2}{HTML}{F8F4F0}
\definecolor{cellc3}{HTML}{F0F8F5}
\definecolor{cellc4}{HTML}{FAF0F5}
\definecolor{cellc5}{HTML}{FDF8E8}
\definecolor{cellc6}{HTML}{F3F0FA}
\definecolor{cellc7}{HTML}{E8F4F8}
\definecolor{cellc8}{HTML}{FDF2F0}
\definecolor{cellc9}{HTML}{EAF7ED}
\definecolor{cellc10}{HTML}{FBF0E8}
\definecolor{cellc11}{HTML}{F5F0FB}
\definecolor{cellc12}{HTML}{EEF6FA}

\definecolor{codegray}{rgb}{0.95,0.95,0.95}
\definecolor{keywordcolor}{rgb}{0.0,0.4,0.8}
\definecolor{stringcolor}{rgb}{0.1,0.5,0.1}

\definecolor{promptbg}{rgb}{0.95, 0.95, 0.95}
\definecolor{promptframe}{rgb}{0.8, 0.8, 0.8}

\lstdefinestyle{aiprompt}{
    backgroundcolor=\color{promptbg},
    frame=single,
    rulecolor=\color{promptframe},
    frameround=tttt,
    breaklines=true,               
    breakatwhitespace=true,        
    basicstyle=\ttfamily\small,    
    columns=flexible,
    keepspaces=true,               
    showstringspaces=false,        
}

\begin{document}

\title[Move Fast and Mend Things]{Move Fast and Mend Things: Keeping Up with Evolving AI Harms Using Social Media Commentary}


\author{Jacqueline Rowe}
\authornote{Research conducted while a Student Researcher at Google.}
\email{}
\orcid{0009-0002-2906-5126}

\affiliation{%
 \institution{The University of Edinburgh}
  \city{Edinburgh}
  \country{UK}
}

\affiliation{%
  \institution{Google}
  \city{London}
  \country{UK}
}

\author{Animesh Srivastava}
\affiliation{%
  \institution{Google}
  \city{}
  \state{California}
  \country{USA}}
\email{}

\author{Sai Teja Peddinti}
\affiliation{%
  \institution{Google}
  \city{}
  \state{California}
  \country{USA}}
\email{}

\author{Seliem El-Sayed}
\affiliation{%
  \institution{Google}
  \city{London}
  \country{UK}}
\email{}

\author{Nina Taft}
\affiliation{%
  \institution{Google}
  \city{}
  \state{California}
  \country{USA}}
\email{}

\renewcommand{\shortauthors}{Rowe et al.}
\newcommand{\nt}[2][]{\todo[author=Nina, color=purple!50, size=\small, #1]{#2}}
\newcommand{\jr}[2][]{\todo[author=Jacquie, color=orange!50, size=\small, #1]{#2}}

\begin{abstract}

The rapid deployment of AI systems has created socio-technical, psychological, and operational harms that can elude ex-ante threat modelling and ex-post incident tracking. We introduce an LLM-assisted thematic analysis pipeline to dynamically detect, categorise, and track emerging AI harms from large-scale social media data. Applying it to 5.7 million Reddit post summaries over 18 months (01/2025 to 06/2026), we curate and release a dataset of 575,000 AI harm-related posts and a bottom-up AI harm taxonomy of 12 categories and 47 subnodes. The taxonomy reliably covers established expert-defined risks while surfacing granular harms that top-down frameworks overlook, such as distinct forms of AI privacy violations. Temporal analysis surfaces evolving user-centric harms, such as agentic privacy and security breaches, premature AI adoption in the workplace, and grief from AI companion discontinuation. Our pipeline shortens harm-detection timelines and hereby complements efforts towards more participatory and responsive AI governance.

\end{abstract}




\maketitle

\section{Introduction}
As AI tools become embedded in workplaces, creative practices, healthcare, and intimate personal interactions, the harms they produce have grown correspondingly diverse and increasingly difficult to anticipate~\cite{unpanel2026preliminary}. Beyond well-documented risks like misinformation, bias, and job displacement~\cite{weidinger2022taxonomy, gabriel2024ethics}, novel capabilities such as persistent conversational AI, autonomous tool use, and deep personalisation are generating emergent harms, from psychological dependency to undetectable security breaches, that existing governance frameworks struggle to address~\cite{brohi2025research, chan2023harms, ghalebikesabi2025privacy, qi2026towards}.

Evaluating adverse consequences of AI requires distinguishing between \textit{AI risk} (the \textit{ex-ante} probability of a hazard occurring) and \textit{AI harm} (materialised setbacks to individuals' interests, autonomy, or lived experience)~\cite{shelby, hansson2013ethics, Oberdiek2014-OBEIRA-2}. Adverse impacts also encompass subjective fears and anticipatory anxieties that constitute significant setbacks even without material harm. While developers implement \textit{ex-ante} measures~\cite{lu2024responsible} like risk assessments, red-teaming and safety alignment measures~\cite{lu2024responsible}, such measures often cannot anticipate harms that emerge in specific downstream contexts, over long periods, or through unpredictable interactions among actors  using AI tools in creative or malicious ways ~\cite{zhang2025dark, chan2023harms, kilian2026beyond, frohling2026agent}. Furthermore, AI risk and policy frameworks tend toward top-down taxonomies informed by controlled studies, rather than capturing how harms are perceived, negotiated and experienced by end-users in practice~\cite{ojewale2025towards, weidinger2025toward, elliott2025accountability}. 

Effective AI risk mitigation therefore requires a grounded understanding of the actualised, adverse lived experiences that emerge from AI deployment over time. HCI and sociotechnical research has addressed this through participatory audits~\citep{deng2025weaudit, devos2022toward, shen2021everyday, lam2022end}, interactive policymaking~\citep{lam2025policy, mun2024particip, kuo2025policycraft}, and pluralistic alignment initiatives~\citep{huang2024collective}, yielding nuanced insights into localised community concerns. However, these methods are resource-intensive, narrow in scope, and slow relative to the speed and scale of current AI deployment, which demands macro-scale approaches to dynamically identify evolving harms.

We explore whether analysing unstructured social media data can offer a distinct approach for identifying AI risks and harms---one that complements both proactive \textit{ex-ante} threat modelling and reactive \textit{ex-post} incident tracking. Treating risk and harm as coupled concepts~\cite{weidinger2022taxonomy}, we seek to understand both anticipatory anxieties about and lived experiences of AI harms, in a way which offers actionable insights for the dynamic AI incident monitoring infrastructure increasingly required by regulators~[\citealp[Art.~72]{eu_ai_act_2024}; \citealp{nist2023ai}], and surfacing where existing safeguards fail in practice. Aware of the epistemological and methodological challenges in large-scale social media analysis, we treat AI-related harm disclosures on social media not as direct ground truths but as partial, distributed and noisy traces of AI harms and concerns in the wild, and critical signals of lived experience vital for reflexive AI governance efforts~\cite{boyd2012critical, olteanu2019social, lupton2016digital, geiger2011trace, marres2017digital}. We structure our research around the following questions:

\begin{description}
    \item[\textbf{RQ1:}] What human-AI interaction harms can we detect from a bottom-up analysis of social media posts about AI, and how do these harms differ from those in existing AI harm taxonomies? 
     \item[\textbf{RQ2:}] How are social media narratives around prominent human-AI interaction harms changing over time, and what does this indicate about changing contexts of AI use?

\end{description}

To answer these questions, we curate a corpus of 5.7 million Reddit posts from 1,000 relevant subreddits between 1 January 2025 and 30 June 2026, and filter it down to $575$k organic, unprompted posts about AI harms. Addressing \textbf{RQ1}, we analyse these posts with an iteratively-developed, human-validated LLM pipeline to construct a bottom-up taxonomy of AI harms and risks, demonstrating that it captures nuanced human-AI interaction harms absent from top-down theoretical frameworks. For \textbf{RQ2}, we assign all posts in our dataset to the newly-created taxonomy and trace how harm narratives evolve over time, identifying both rapid spikes tied to specific platform events and steady growth in concerns such as agentic privacy and cybersecurity risks, premature workplace AI adoption, and grief from AI companion discontinuation.  

\paragraph{Contributions and Relevance} 
This work sits at the intersection of critical data studies, algorithmic governance, and system design. We explore how to `move fast' in detecting and addressing emerging AI harms using online discourse, which can directly inform responsive AI design and targeted HCI research. We seek to advance the methodological and substantive goals of human-centred AI safety via three primary contributions:

\begin{itemize}
    \item \textbf{Methodological Pipeline:} A reusable, LLM-assisted qualitative annotation architecture reflexively designed to construct taxonomies of AI-related risks and harms from organic text data about human-AI interactions, facilitating real-time risk monitoring.
    \item \textbf{Curated Dataset and Taxonomy of AI Harms from Reddit}: An expert-curated, publicly accessible dataset of $575$k LLM-annotated Reddit posts discussing AI harms from January 2025 to June 2026, paired with a bottom-up taxonomy that serves as a diagnostic tool for surfacing granular, vernacular user harms and blind spots overlooked by prior top-down frameworks. 
    \item \textbf{Empirical Insights:} Qualitative and quantitative analyses of Reddit users' perceptions of existing and emerging AI-related harms, demonstrating how social media discussions about people's experiences and anxieties on these topics can inform concrete recommendations for AI regulators, researchers and developers.
\end{itemize}

\section{Related Work}
\label{sec:related_work}

\subsection{Mapping AI Risks and Harms}
\label{sec:related_work_1}
A wide range of interventions seek to proactively map AI risks and mitigate harms, spanning \textit{ex-ante} measures such as risk assessments~\cite{nist2023ai}, safety benchmarking~\cite{liang2022holistic, ren2024safetywashing, vanschoren2025role}, red-teaming~\cite{ganguli2022red}, and user consultations~\cite{delgado2023participatory}, alongside \textit{ex-post} measures including incident tracking, independent audits, and algorithmic impact assessments~\cite{lam2022end, brundage2020toward, raji2020closing}. \textit{Ex-ante} risk assessments are now mandated by the EU AI Act for high-risk systems~\cite[Arts.~9, 55]{eu_ai_act_2024} and encouraged by US Department of Commerce's National Institute of Standards and Technology (NIST)~\cite{nist2023ai}. \textit{Ex-post} incident databases are also increasingly popular tools. The Organisation for Economic Co-operation and Development (OECD)'s AI Incidents and Hazards Monitor~\cite{oecd_aim_2026} uses LLMs to catalogue over $17{,}000$ incidents from international news outlets, while the Responsible AI Collaborative's AI Incident Database~\cite{mcgregor2021preventing} crowdsources over $1{,}600$ reports from news, industry, and academic sources.

Taxonomies of AI risks and harms serve as vital scaffolding for risk mitigation, helping practitioners systematically categorise complex risk landscapes and decide upon operational priorities. Existing AI risk taxonomies use distinct foundational inputs and serve different analytical aims. For instance, IBM's AI Risk Atlas~\cite{bagehorn2025ai} takes an engineering perspective across the model lifecycle (covering 23 training, inference, output, and non-technical risks across generative and agentic AI). In contrast, interdisciplinary frameworks such as \citet{weidinger2022taxonomy} (21 LLM risks) and the MIT AI Risk Repository~\cite{slattery2026ai} (24 subdomains) organize harms across broad societal domains---including \textit{Discrimination and Toxicity, Privacy and Security, Misinformation, Malicious Use, Human-Computer Interaction, Socioeconomic and Environmental Harms} and \textit{Systemic and Existential Risks}. These taxonomies are evidence-backed, expert-informed academic tools that structure AI risk across diverse domains. Yet any approach targeted at addressing novel or changing AI risks and harms requires significant speed and flexibility. AI risk and harm research tends to lag behind the pace of AI innovation, development, and deployment, resulting in a reactive approach to mitigating harm rather than proactive measures that minimise risks and prevents harms in the first place~\cite{kashefi2024shaping, taeihagh2021governance, currie2025rethinking}. AI model and tool capabilities are constantly adapting, and their deployment across novel social contexts triggers changes in users' behaviour over time, continually altering the AI harms landscape in often unpredictable ways~\cite{selbst2019fairness, devos2022toward, oneill2022contextual}. Expert-informed and literature-derived taxonomies require resource-intensive consultation processes to create and periodically update, and their reliance on academic literature, expert perspectives and media reports may not capture harms in time or as described vernacularly by users and affected parties, who hold different epistemological standpoints and experiences of AI systems~\cite{BIRHANE2021100205, gautam2025towards}. There is a clear need to complement these efforts with more dynamic risk monitoring processes integrating a greater variety of perspectives to facilitate more responsive AI harm mitigation efforts~\cite{nist2023ai, lu2024responsible}.

\subsection{LLM-Facilitated Analysis of AI Harms Discourse on Social Media}
\label{sec:related_work_2}

Social media analysis is a widely-used approach for gathering large-scale insights into public perceptions of technological systems ~\cite{praveen2022general, okey2023investigating, rao2025quallm, zhou2026ai, xu2024public, koonchanok2024public, thaiduong2026public, thomas2026understanding}, and can capture unprompted, emergent themes that structured surveys or targeted consultations may miss~\cite{raghupathi2026exploring, grassini2026potential}. Social media analysis can also capture perspectives on AI-related harms that product and complaint logs may miss, as AI impacts extend far beyond direct users of a system~\citep{shelby} and certain structural worries or complaints \textit{about} AI tools, such as fears of unauthorized data use, are not discernable from system telemetry~\citep{shen2021everyday, raji2022outsider}. Existing large-scale social media analyses of AI-related harms often focus on specific product categories~\cite{rao2025quallm, koonchanok2024public, li2025towards}, potentially missing broader, cross-system harms that must be addressed by coordinated AI governance efforts. As a result, we define AI-related harms more broadly and include online perspectives on diverse ecosystems of AI tools.

While social media data allows us to investigate perceptions of AI harms across a large number of people and communities, it carries inherent biases. Disclosures in specific online communities are not representative of offline populations, and may not reliably align with users' actual perspectives and behaviours~\cite{proferes2021studying, alvarado2025what}. Social media data is often selected for its availability and volume rather than its contextual appropriacy~\cite{boyd2012critical}, and platforms increasingly contain AI-generated content alongside organic content~\cite{sun-etal-2025-ai}. Furthermore, traditional methods for computational text analyses---such as topic modelling, sentiment analysis, and text summarisation~\cite{zhou2026ai, xu2024public, koonchanok2024public}---often impose misleading claims to objectivity without critical reflection on how they flatten ambiguities and overlook context~\cite{boyd2012critical, alvarado2025what}. We do not position online posts as direct reflections of lived reality, nor large-scale social media analysis as a substitute for participatory AI harms research. Instead, we view this data as \textit{one of many }valuable lenses on human-AI interaction harms, which carries the possibility of providing early-warning signals for emerging risks that take time to investigate through complementary research methods.

Recent work posits that LLMs can combine the reflexivity and interpretative depth of inductive coding and thematic analysis with the scalability of NLP tools for large-scale text analysis~\cite{rao2025quallm, randerson2026exploring, wang2025lata}, with LLMs now able to match or outperform human annotators on data labelling tasks~\cite{gilardi2023chatgpt, alizadeh2023open, tornberg2023chatgpt4} or thematic analysis based on existing codebooks~\cite{chew2023llm, randerson2026exploring}. However, uncritical usage of LLMs for textual analysis raises valid concerns about bias, inconsistency, and limited transparency over research findings~\cite{ashwin2026using, depaoli2024performing, wang2025lata, randerson2026exploring, shen2023shaping, pezeshkpour2024large, zheng2024large}. As discussed in detail below (\Cref{sec:method}), we mitigate these risks by taking a reflexive and iterative approach, conducting several rounds of data exploration and manual analysis to build thematic codebooks and evaluation data at each stage of data processing to ensure high confidence in the findings.

\section{Method}
\label{sec:method}

We design an LLM-assisted pipeline to automatically generate AI harm taxonomies from organic social media discourse. While the pipeline is data-agnostic, we apply it to Reddit for its scale ($450$M weekly active users from $180$ countries~\cite{worldpop2026reddit} and over $100{,}000$ active subreddits~\cite{reddit2026form}), anonymity, and permissive content policies that encourage candid disclosure~\cite{proferes2021studying, raghupathi2026exploring}. We first collect $575\text{k}$ posts about AI harms from $1{,}000$ relevant subreddits across 18 months of data (\Cref{sec:post_collection}). We then develop a thematic codebook and apply it to each post using an LLM (\Cref{sec:thematic_analysis}). Finally, we construct a hierarchical taxonomy by prompting an LLM to extract and iteratively merge key topics from post summaries, and assign all posts to taxonomy nodes (\Cref{sec:taxonomy_gen}). We use Gemini 3.5 Flash Lite throughout for its performance and cost-effectiveness (Appendix~\ref{sec:appendix_params} and Appendix~\ref{sec:appendix_compute_cost}), and report LLM accuracy against manually-labelled data with Inter-Annotator Agreement (IAA) scores at each stage (Appendix~\ref{sec:appendix_iaa}). All prompts are included in the supplementary materials, and we release our LLM-annotated and paraphrased dataset as a public research tool.\footnote{Following CHI publication guidelines, the anonymised dataset repository will be hosted on an open-access platform (e.g., Open Science Framework / re3data.org) upon acceptance.}

\paragraph{\textbf{Ethics Statement}}
We acknowledge that the public availability and permissive licensing of Reddit posts does not equate to implicit user consent for research reuse~\cite{boyd2012critical, fiesler2024remember, fiesler2018participant}. Our approach aligns with consensus among Institutional Review Boards that human subjects review is not required for publicly available and pseudo-anonymous data which does not include personally-identifiable information~\cite{proferes2021studying,li2025towards}. We mitigate potential privacy risks by fully anonymising and paraphrasing all qualitative examples cited in this paper and including only post \textit{summaries} -- not the original Reddit posts -- in the public version of our dataset. This prevents searchability of Reddit posts based on our research and aligns with Reddit's Terms of Service. We also use an LLM API which guarantees that input data is never ingested by model providers, ensuring that Reddit posts are not consumed as training data as a result of our work. Additionally, we recognise that Reddit overrepresents younger, male, American and highly-educated users~\cite{proferes2021studying,worldpop2026reddit, furrh2026reddit}. As such, we do not present our Reddit-based insights as representative of broader public concern about AI risks and harms, and ensure that our pipeline can be easily applied to other sources of online discourse to encourage future work centring other perspectives. 

\subsection{Collecting AI harms posts from Reddit}
\label{sec:post_collection}
For this research, we follow an approach similar to \citet{thomas2026understanding}. We source data from a pre-existing policy-compliant web crawl of Reddit (respecting its \texttt{robots.txt}), which continuously tracks public Reddit URLs and actively handles post deletions respecting user privacy.\footnote{This repository is managed by an organization that some co-authors are affiliated with, and will be cited in full upon acceptance.} As this dataset has only partial visibility into comments, we omit all comments from our analysis. We target posts describing direct or indirect negative concerns or materialised outcomes stemming from the use or deployment of AI technologies, broadly defined as those leveraging machine learning, LLMs, automated decision-making algorithms, or generative media.

To find these posts at scale, we first collect and manually label a balanced dataset of AI-harm and non-AI-harm posts ($N=124$) from 90 manually-identified subreddits which contain discussions about AI harms (Appendix~\ref{sec:appendix_subreddits}). We discuss and refine our definition of AI harms through analysing examples and edge cases from this sample; for example, we exclude posts generically discussing `bots' without explicit reference to an AI capability, and include hypothetical discussions of AI harms \textit{only} if they clearly express psychological harms such as anxiety, dread or distress. We then iteratively develop and evaluate an AI harm filtering prompt using this seed dataset.

Next, we apply this prompt to a stratified sample of posts from \textit{all} active subreddits across a 60-day period (1 May--30 June 2026), removing URLs and filtering out posts shorter than 50 words and subreddits with fewer than 60 remaining posts across the time period. We select the 1,000 subreddits with the highest proportions of AI harm posts and create a second manually-labelled seed dataset randomly sampled from these channels ($N=310$), which we manually label and use to further refine the filtering prompt. For example, noticing a particular category of false positives using one instantiation of our prompt, we edited the instruction to ensure that posts that do not explicitly reference an AI tool or that contain only generic complaints about account access or corporate policies are excluded. Finally, we collect 5.7 million posts from these subreddits between 1 January 2025 and 30 June 2026, apply a 50-word minimum length threshold, and use our final filtering prompt to identify $575\text{k}$ AI harm-related posts.

\paragraph{Evaluation} We validated the LLM's final post filtering accuracy by manually reviewing a balanced random sample of 100 posts from the 18-month dataset, comprising 50 predicted positive (`AI harm') and 50 predicted negative (`non-AI harm') cases. Two researchers independently evaluated the LLM-assigned labels as correct or incorrect (IAA reported in Appendix~\ref{sec:appendix_iaa}), and then resolved disagreements via consensus. According to these labels, the LLM achieved an accuracy of $0.84$ and an F1 score of $0.81$ (precision = $0.70$, recall = $0.97$). We prioritise high recall during this stage to avoid discarding relevant but non-standard expressions of AI harms, relying on downstream taxonomy assignment to filter out irrelevant noise and false positives collected using this prompt.

\subsection{LLM-facilitated Thematic Analysis and Text synthesis}
\label{sec:thematic_analysis}
To extract useful insights about AI harms and risks from our dataset, where posts are often noisy and include unrelated topics, we use an LLM to extract key information from each post and generate a concise summary of the AI harm described. We develop a thematic codebook drawing on existing codebooks for large-scale social media analyses~\cite{thomas2026understanding, rao2025quallm}, prior categorisations of AI harms~\cite{oecd_aim_2026, mcgregor2021preventing}, and manual exploration of sample posts. We refine the codebook by applying it to a random sample of 200 posts from our dataset, inspecting how labels are assigned and investigating misclassifications, and editing the prompt by clarifying definitions or adding codes as required. We repeat this iterative process three times. 

The final codebook consists of six features. Two focus on who is involved in the AI-related harm being described; (i) the entity affected by the harm, and (ii) the entity using the AI tool, with options ranging from \textit{first-person author} to \textit{people in general}. The next two features infer information about the text author's perspective of the AI-related harm being described, extracting (iii) the apparent objective of the author in making the post -- for example, \textit{raising awareness}, \textit{seeking advice}, or \textit{inspiring debate} -- and (iv) how severe they seem to consider the harm in question (\textit{high, moderate} or \textit{low}). Finally, two features collect information about the possible underlying causes of the AI-related harm, focusing on (v) any specific datatypes that are referenced in the harm description, such as \textit{illegal content, personally identifiable information} or \textit{synthetic data}; and (vi) any specific AI capabilities that are connected with the harm being described, such as \textit{vision processing, search or information retrieval}, or \textit{digital or physical actuation}. We prompt the LLM to apply this codebook to all $575$k posts and then generate a 2--3 sentence summary of each post using both the original text and the assigned codes.

\paragraph{Evaluation} To evaluate the LLM's thematic coding reliability, two researchers independently reviewed model-assigned codes across a stratified sample of 94 posts (two posts per supported category across the six features), rating each model label as correct or incorrect before resolving discrepancies through consensus. The LLM demonstrated high reliability across most dimensions, achieving between 80\% and 100\% precision across five of the six features (see Appendix~\ref{sec:appendix_iaa} for inter-annotator agreement and per-feature breakdowns). For the lone outlier -- \textit{Entity Using AI tool} (67\% precision) -- inter-annotator agreement was correspondingly lowest, reflecting the fact that this information was often difficult to extract correctly given that many posts are vague with multiple valid interpretations for this feature, even by human annotators. We retain this feature in our pipeline due to its exploratory utility in data analysis downstream, but highlight possibilities for further improving LLM code assignment through using certainty thresholds to minimise errors in future work. 

Second, we evaluate Gemini's text synthesis performance by adapting~\citet{rao2025quallm}'s evaluation framework for LLM-generated summaries of concerns articulated in Reddit posts. Two researchers manually annotate the summaries generated for 100 randomly-sampled posts, giving a score of 0 or 1 for factuality (\textit{`Is the summarised AI harm described in the LLM text present in the original Reddit post?'}) and completeness (\textit{`Is the primary AI harm described in the original Reddit post included in the LLM summary?'}) (IAA metrics in Appendix~\ref{sec:appendix_iaa}). After resolution to obtain ground truth labels, the LLM summaries are judged as factual in 93\% of cases, and as complete in 96\% of cases. 

\subsection{Taxonomy Generation and Assignment}
\label{sec:taxonomy_gen}
Unlike our deductive coding in Section~\ref{sec:thematic_analysis}, we construct our taxonomy entirely bottom-up without seeding or anchoring on existing AI harm taxonomies or prior research, ensuring the hierarchy surfaces organic, user-defined harms directly informed by the themes discussed in the posts. Established approaches for generating taxonomies from unstructured text broadly include clustering-based methods (e.g., TaxoGen~\cite{taxogen}) and LLM-based methods (e.g., TnT-LLM~\cite{tnt-llm}, TaxoAdapt~\cite{taxoadapt}). However, these methods carry limitations; the former produces keyword clusters that lack canonical naming, formal definitions, and clear semantic boundaries and the latter relies heavily on data downsampling due to context-window constraints, forfeiting coverage over our large-scale Reddit dataset.

To combine the rich semantic reasoning capabilities of frontier LLMs with dataset-scale coverage, we introduce a scalable three-stage taxonomy generation pipeline. We first prompt the LLM to extract concise, semantically-grounded concepts from each Reddit post summary (e.g., ``\textit{API token leaks}'', ``\textit{sycophancy and excessive agreement}'', and ``\textit{black box user experience}''), yielding an average of $4.89$ concepts per post and $1.6$ million concepts across the entire dataset of $575$k posts. Second, we progressively consolidate these concepts by prompting the LLM to combine batches of $200$ concepts at a time into a merged list, eliminating duplicate concepts and lexical divergence. We repeat this process twice, resulting in a list of $43$k merged concepts. Finally, we rank these merged concepts by post prevalence (number of posts whose raw concepts were subsequently mapped into the merged concept) and pass the top $20{,}000$ in a single prompt to the LLM, instructing it to organise them into a `granular, comprehensive and actionable' two-level hierarchy based on natural semantic clusters. We specify that each parent node should contain between two and eight subnodes, and that the subnode definitions should include a description of: (i)~\textit{Mechanisms} (underlying technical failures or triggering conditions), (ii)~\textit{Impact} (the user experience and the severity of the harm), and (iii)~\textit{Scope Boundaries} (criteria delineating it from adjacent categories). This three-stage approach balances comprehensive, dataset-wide coverage with context window constraints, generating clearly delineated, formally defined harm categories rooted in salient themes from the dataset. 

For Taxonomy Assignment, we instruct the LLM to analyse the text of each full Reddit post, and to select one or more appropriate taxonomical categories based on its semantic understanding of the text. Following the prompting strategy of~\citet{simula_tmlr}, within each L1 theme, we prompt the LLM to consider whether each subnode directly and specifically describes a core aspect of the text, assigning the label where this is true. Where no subnode is assigned, we also prompt the LLM to evaluate whether the parent node category applies directly. We permit multi-label assignments for posts which clearly meet multiple criteria, but restrict assignments to high-confidence ($\ge 70\%$ certainty).

\paragraph{Evaluation of Taxonomy Generation}
Our bottom-up taxonomy is designed to function as a heuristic tool for identifying existing and emerging AI harms as described by users, including perspectives and themes which may not be adequately represented in top-down expert frameworks. To evaluate this diagnostic utility, we compare our generated taxonomy to an expert-curated taxonomy, choosing the MIT AI Risk taxonomy~\cite{slattery2026ai} due to its inclusion of a range of human-centric AI risks such as downstream social and psychological harms (see \Cref{sec:related_work_1}). Two researchers independently coded each MIT node as \textit{covered}, \textit{partially covered}, or \textit{not covered} based on node names and descriptions, resolving disagreements through consensus (IAA scores in Appendix~\ref{sec:appendix_iaa}). Our taxonomy fully covers $13$ MIT nodes, partially covers $8$, and does not cover $3$ (Appendix~\ref{appendix:mit_comparison}). Nodes not covered are abstract, societal-level concerns such as \textit{AI welfare and rights} and \textit{Power centralisation}. In `partially-covered' nodes, the MIT taxonomy includes broader characterisations of harms than those reflected in our taxonomy; for instance, ours captures cyberattacks but does not explicitly mention weapons development, and system and platform transparency deficits but not model interpretability issues. For these partially and not covered topics, upon investigation we find that our dataset does include some posts relevant to these themes, but they make up too small a proportion of our overall dataset to merit their own taxonomy node, reflecting Reddit's orientation toward individual, everyday interactions. Conversely, our taxonomy captures detailed, vernacularly-described harms not present in the MIT taxonomy, such as algorithmic disparities specifically impacting gig workers, in-product training data consent violations, operational burdens associated with premature AI adoption in the workplace and grief from AI companion discontinuation---which we explore in-depth in Section~\ref{sec:results}. This mapping demonstrates that our bottom-up method reliably surfaces all well-documented, high-coverage human-AI interaction harms.

\paragraph{Evaluation of Taxonomy Assignment}
We validate the model's taxonomy assignment by having two researchers independently evaluate the LLM labels for a diverse and random sample of $60$ posts with 129 taxonomy labels (ensuring every node or subnode label had a representation). They checked whether `All'/`Some'/`None' of the assigned labels for each post were correct before resolving discrepancies through consensus. The inherent complexity of this task of categorising Reddit posts into appropriate AI harm category labels is underscored by the relatively low human agreement scores, with $P_o = 71.7\%$ and Gwet's $\text{AC}_1 = 0.645$ (see Appendix~\ref{sec:appendix_iaa}). These modest agreement scores stem from two factors: first, the raw Reddit posts are often noisy and ambiguous, permitting multiple plausible interpretations; and second, because AI harms are themselves complex sociotechnical phenomena with fluid semantic and contextual boundaries. Against the consensus ground-truth, we observe that all the 60 posts had `Some'  or `All' labels correct, with the zero-shot LLM achieving a precision of 89.2\% on the diverse sample across 59 possible labels (over $52\times$ the uniform random baseline chance of $1.69$\%). 
Future work could explore hierarchical top-down routing, greater certainty thresholds, or few-shot prompting to further optimize taxonomy assignment precision. We proceed with these assignments for our downstream analyses. 

\section{Results}
\label{sec:results}
This section discusses our findings in relation to our two research questions, based on our taxonomy and how posts are assigned to it. We present our bottom-up taxonomy of AI risks and harms in \Cref{tab:18_month_taxo_short}, with the complete definitions of each node and subnode in Appendix~\ref{sec:appendix_full_taxonomy} and the distributions of post assignments across all nodes and subnodes in Appendix~\ref{appendix:taxonomy_distribution}.

\begin{table*}[t]
\centering
\small
\begin{tabularx}{\textwidth}{XXX}
\toprule
\cellcolor{cellc1}
\textbf{AI Hallucinations, Fabrications, and Inaccuracies} 
\footnotesize{
\begin{itemize}[leftmargin=*, noitemsep, topsep=0pt, parsep=0pt]
\item Calculation and Quantitative Processing Errors
\item Code Generation Technical Errors and Logic Flaws
\item Factual Hallucinations and Confabulations 
\item Instruction Following and Prompt Adherence Failures
\end{itemize}}
&
\cellcolor{cellc2}
\textbf{Algorithmic Bias, Discrimination, and Unfairness} 
\footnotesize{
\begin{itemize}[leftmargin=*, noitemsep, topsep=0pt, parsep=0pt]
\item Algorithmic Discrimination and Demographic Disparity 
\item Automated Hiring, Recruitment, and Evaluation Bias
\item Gig Economy Algorithmic Management Injustices
\item Recommender System Manipulation and Bias
\end{itemize}}
&
\cellcolor{cellc4}
\textbf{Autonomous Agent, Tool Execution and Digital Actuation Risks} 
\footnotesize{
\begin{itemize}[leftmargin=*, noitemsep, topsep=0pt, parsep=0pt]
\item Autonomous Vehicle and Navigation Safety Failures 
\item Runaway Autonomous Agent Actions and Infinite Loops 
\item Smart Device, IoT, and Robotics Malfunctions 
\item Tool-Calling and Function Execution Errors
\end{itemize}}
\\
\midrule
\cellcolor{cellc3}
\textbf{Cybersecurity Vulnerabilities and Attack Vectors}
\footnotesize{
\begin{itemize}[leftmargin=*, noitemsep, topsep=0pt, parsep=0pt]
\item AI-Assisted Malware and Cyberattack Generation 
\item API Key Compromise and Credential Leakage 
\item Prompt Injection and Jailbreaking Exploits 
\item Supply Chain and Third-Party Integration Vulnerabilities 
\end{itemize}}
& 
\cellcolor{cellc5}
\textbf{Financial Harm, Scams, and Economic Burdens}
\footnotesize{
\begin{itemize}[leftmargin=*, noitemsep, topsep=0pt, parsep=0pt]
\item Financial Fraud, Scams, and Impersonation 
\item Macroeconomic Job Displacement and Livelihood Loss
\item Predatory Lending and Automated Debt Collection Harassment 
\item Unexpected Billing Spikes and Financial Overcharges
\end{itemize}}
&
\cellcolor{cellc6}
\textbf{Intellectual Property Infringement and Copyright Violations} 
\footnotesize{
\begin{itemize}[leftmargin=*, noitemsep, topsep=0pt, parsep=0pt]
\item Artistic Style Mimicry and Unauthorized Creative Reproduction 
\item Copyrighted Code and Proprietary Data Ingestion 
\item Plagiarism and Attribution Failures in Academic and Professional Work 
\end{itemize}}
\\
\midrule
\cellcolor{cellc7}
\textbf{Misleading Content, Media Manipulation, and Deception} 
\footnotesize{
\begin{itemize}[leftmargin=*, noitemsep, topsep=0pt, parsep=0pt]
\item Deepfake and Synthetic Media Abuse 
\item Misleading AI Capabilities and Vendor Hype 
\item Political Disinformation and Propaganda Proliferation 
\item Unlabelled AI Content Flooding and Slop
\end{itemize}}
&
\cellcolor{cellc8}
\textbf{Platform Governance, Moderation Overreach, and Redress Deficits} 
\footnotesize{
\begin{itemize}[leftmargin=*, noitemsep, topsep=0pt, parsep=0pt]
\item Absence of Human Recourse, Appeal, and Customer Support
\item False-Positive Content Moderation and Over-Censorship 
\item Unilateral Service Modifications and Policy Changes
\item Wrongful Account Bans and Access Revocation
\end{itemize}}
&
\cellcolor{cellc9}
\textbf{Privacy Violations and Data Abuses} 
\footnotesize{
\begin{itemize}[leftmargin=*, noitemsep, topsep=0pt, parsep=0pt]
\item Personally Identifiable Information (PII) Exposure and Leaks 
\item Commercially Sensitive Data and Code Leakage 
\item Model Training Data Consent Violations 
\item Surveillance, Tracking, and Behavioural Monitoring 
\end{itemize}} 
\\
\midrule
\cellcolor{cellc10}
\textbf{Psychological Distress and Emotional Harm} 
\footnotesize{
\begin{itemize}[leftmargin=*, noitemsep, topsep=0pt, parsep=0pt]
\item AI Addiction, Dependency, and Over-reliance 
\item AI-Induced Anxiety and Existential Distress 
\item Grief from AI Companion Discontinuation or Persona Drift
\item Severe Mental Health Crises and Self-Harm Facilitation
\end{itemize}}
&
\cellcolor{cellc11}
\textbf{Technical Glitches, System Failures, and Outages}
\footnotesize{
\begin{itemize}[leftmargin=*, noitemsep, topsep=0pt, parsep=0pt]
\item Context Window Exhaustion and State Loss 
\item Service Outages and Infrastructure Downtime 
\item Software Bugs, Crashes, and Memory Leaks 
\item Software Update Breakage and Backward Incompatibility
\end{itemize}}
&
\cellcolor{cellc12}
\textbf{User Frustration and Dissatisfaction} 
\footnotesize{
\begin{itemize}[leftmargin=*, noitemsep, topsep=0pt, parsep=0pt]
\item General User Frustration and Abandonment
\item Low-Severity Annoyances and Minor Inconveniences 
\item Subscription and Paywall Friction 
\item Workflow Friction and Operational Disruption 
\end{itemize}}
\\
\bottomrule
\end{tabularx}
\caption{Nodes and subnodes of our taxonomy of AI harms derived from 18 months of Reddit posts. See Appendix~\ref{sec:appendix_full_taxonomy} for node and subnode descriptions, which include boundary conditions for each category.}
\label{tab:18_month_taxo_short}
\end{table*}

\subsection{RQ1: What human-AI interaction harms can we detect from a bottom-up analysis of social media posts about AI, and how are these different from those in existing AI harm taxonomies?}
\label{sec:results_RQ1}
\textit{In summary, this bottom-up taxonomy surfaces broad categories of well-known AI harms, and additionally reveals more granular and user-centric harms that traditional frameworks miss. As such, it reflects specific hotspots of novel concerns \textbf{within} areas such as algorithmic bias, false or misleading content, or privacy violations}. Our taxonomy covers twelve broad AI harm nodes (L1s) with 47 subnodes (L2s), which can be grouped into four categories. First, some of the L1 nodes align with classic technology-facilitated harms, such as scams, cybersecurity issues and privacy violations. A second category refers to social-media and platform-related harms that are being exacerbated by the widespread adoption of AI tools, such as media manipulation, algorithmic bias, and platform governance issues. Third, several L1 nodes capture risks and harms unique to novel AI training paradigms and deployments. These include hallucinations, autonomous agent risks and copyright infringements, alongside emerging contexts of AI usage which contribute to psychological harms. Finally, two L1 nodes -- technical glitches and user frustrations -- capture operational friction, due to the fact that we purposefully employed a broad, general prompt when filtering our initial corpus for AI-related harms and prioritised recall over precision in order to increase the chances of catching emerging harms. User frustrations is by far the most frequently assigned category in the dataset distribution (45\% of posts), followed by \textit{AI Hallucinations, Fabrications and Inaccuracies} (16\%) and \textit{Psychological Distress and Emotional Harm} (14\%). The volume of user frustration posts is expected due to the fact we allow for multi-node assignment; users describing other harms in our taxonomy are often simultaneously experiencing frustration, resulting in this tag being commonly applicable in addition to whatever other harm is being described. However, as many of the posts in this category describe routine usability complaints that are addressable through standard product improvement processes rather than novel policy or governance interventions, they are likely to be lower priority for policymakers and researchers focused on emerging AI-specific harms.

While our L1 nodes capture broad, widely-recognised dimensions of AI harm, the L2 nodes reveal specific themes within those areas which are particularly prevalent in Reddit discussions during our dataset period, providing useful nuance for AI developers, researchers and policymakers alike. While a full qualitative exploration of every subnode in the taxonomy is beyond the scope of this paper, here we highlight three harm areas where we found the detail included in subnodes particularly insightful; algorithmic bias; false or misleading content; and privacy harms. Additional quotes relevant to each of these areas are included in Appendix \ref{appendix:privacy_examples}.

\subsubsection{Algorithmic Bias, Discrimination and Unfairness} Early scholarship in algorithmic fairness and transparency highlights the relationship between model performance disparities and broader socio-technical inequalities~\citep{benjamin2019race, selbst2019fairness, hoffmann2019fairness}. Subnodes within this node in our taxonomy surface specific domains of lived experiences of algorithmic fairness harms amongst Reddit users. The most populous subnode is \textit{\textbf{Gig Economy Algorithmic Management Injustices}}, where gig workers frequently post in subreddits dedicated to delivery drivers or paid data labellers to highlight perceived disparities in pay, and what they describe as \textit{`deceptive'}, \textit{`manipulative'} and \textit{`unpredictable'} remuneration algorithms which contribute to economic harms and inequalities. A separate subnode entitled \textit{\textbf{Automated Hiring, Recruitment and Evaluation Bias}} focuses instead on concerns around how \textit{`dehumanising'} AI tools -- such as \textit{`one-way AI interview software'} -- are being used for scoring, selecting or rejecting potential job candidates, with little to no transparency about their judging criteria nor appropriate accommodations for people with disabilities. These hotspots of concern \textit{within} the algorithmic bias category indicate AI-related fairness harms that directly impact Reddit users' day-to-day lives and economic stability.

\subsubsection{False and misleading content} Hallucinations are commonly described as a monolithic category in AI safety research, yet we find that Reddit users treat hallucinations in different ways. For instance, our \textit{\textbf{AI Hallucinations, Fabrications and Inaccuracies}} node differentiates model failures with distinct downstream consequences. Subnodes include instruction-following errors, factual hallucinations, code generation flaws and quantitative and calculation inaccuracies. Users describe impacts of these errors ranging from loss of productivity to incorrect tax advice to epistemic mistrust. Beyond these nuanced subnodes, another L1 node covering hallucinations is \textit{\textbf{Misleading Content, Media Manipulation, and Deception}}. This node focuses on \textit{deliberately} falsified content rather than accidental errors. For example, the most populous subnode is \textit{\textbf{Deepfake and Synthetic Media Abuse}}, which primarily includes posts related to deepfake non-consensual intimate imagery. Some authors report that they were targeted with \textit{‘revenge porn’} after a widely-used AI tool released a specific feature that allowed users to manipulate photos to remove people's clothing and post them publicly. Some Reddit users also discussed the adverse impact that such tools have on those struggling with porn addiction. Together, these subnodes illustrate how false or manipulative AI outputs span a continuum from operational friction and confusing misinformation to severe psychological, reputational, and bodily harms. They also highlight the range of reasons underlying different harms ranging from model misalignment to misuse by adversarial actors.

\subsubsection{Privacy} Finally, AI privacy harms -- which are often described as a monolithic category in top-down taxonomies -- represent 74,000 posts across four distinct subnodes. The \textbf{\textit{PII Exposure and Leaks}} subnode frequently captures individuals' fears that private inputs to AI tools may be leaked or made public, even after account or conversation deletion. For example, one Reddit post author had been using a chatbot as a therapist to discuss intrusive thoughts. While they had deleted all chat history and their account, they said they \textit{``regret''} sharing this information and they were \textit{``terrified I'll be flagged by the police or my chats will be leaked,''} even though they had not done anything illegal. Others posts in this subnode report concerns about third parties -- such as medical professionals -- inputting personal or sensitive data into AI tools without their consent, as well as uncertainty around providing AI agents with sensitive data like credit card details. The second L2 privacy node, {\textbf{\textit{Commercially Sensitive Data and Code Leakage}}}, categorises harms involving cloud-based AI tools exposing trade secrets and proprietary code, or confidential business information. Some reports centre on technical errors; one user reportedly observed an AI tool randomly \textit{``including API keys, internal URLs and a production dashboard endpoint ...  pulling context from a completely different user’s
environment''}. Other discussions relate to individuals erroneously sharing commercially sensitive data inappropriately with AI tools. For example, one Reddit post author describes witnessing a junior colleague copying and pasting\textit{ ``200+ customer records including phone numbers, emails, and purchase histories''} into an online Chatbot while debugging a database query, prompting an overhaul in company policy and security governing the use of AI tools. 

Our taxonomy also includes \textbf{\textit{Model Training Data Consent Violations}} as a third privacy subnode, which captures profound anxieties regarding data harvesting users are not empowered to control. Reddit users frequently express dismay at discovering their data is being used to train AI models, reporting feelings of \textit{``horror''} and \textit{``invasions of privacy''}. Many post authors also report that they find it difficult to opt-out of such procedures, whether because settings reset with updates or because the process is unnecessarily opaque; some discuss defensive tactics such as seeking \textit{``anti-AI filters''} to protect personal artifacts from ingestion into training data pools. The final privacy subnode, \textbf{\textit{Surveillance, Tracking, and Behavioural Monitoring}}, surfaces examples of how AI increases the possibilities for surveillance -- such as ambient tracking embedded in AI-driven software having a severe chilling effects on users. Some posts in this category describe a perceived function creep of AI-enabled smart home devices towards more invasive personalisation. For example, one user reports that their smart home device has \textit{``got a lot creepier''}, increasingly detecting and commenting on emotions in their voice or their music choices, as well as bringing up \textit{``things I said before''}. Posts in this subnode also describe significant workarounds that Reddit users adopt to combat digital surveillance, such as quitting jobs where employers use AI monitoring software, or strictly running integrated AI tools in sandboxed or private browser tabs to prevent tech companies from linking their usage data across different applications. 

\subsection{RQ2: How are social media narratives around prominent human-AI interaction harms changing over time, and what does this indicate about changing contexts of AI use?}
\label{sec:whats_changing}
\textit{In summary, social media narratives about AI harms mirror evolving AI capabilities, contexts of deployment and normative expectations, with temporal analysis revealing both gradual shifts in frequency of posts over time as well as event-driven spikes tied to specific model releases or updates.}
The monthly frequency of posts in our dataset increases throughout the 18-month period, indicating both accelerating mainstream adoption and awareness of AI tools and a proportional rise in the harms users describe and discuss on Reddit. To examine how Reddit narratives about human-AI interaction harms shift over time, we investigate harm categories whose volume grows disproportionately relative to overall dataset growth. We present three illustrative case studies with distinct growth patterns: (i) a rapid acceleration of concern and anxiety about agentic privacy and security; (ii) steady growth in reported challenges surrounding premature workplace AI adoption; and (iii) episodic spikes in posts describing grief from AI companion discontinuation. While our dataset contains many other compelling trajectories of evolving harms, we chose these three to illustrate diverse temporal modes of change and to demonstrate the deep granularity of insights our approach can surface. For each case study, we integrate temporal and qualitative analyses to identify sociotechnical and contextual factors driving these evolving narratives.

\subsubsection{The rise of agentic privacy and security concerns} 
\label{sec:agents}

\begin{figure*}[h]
    \centering
    \includegraphics[width=1\linewidth]{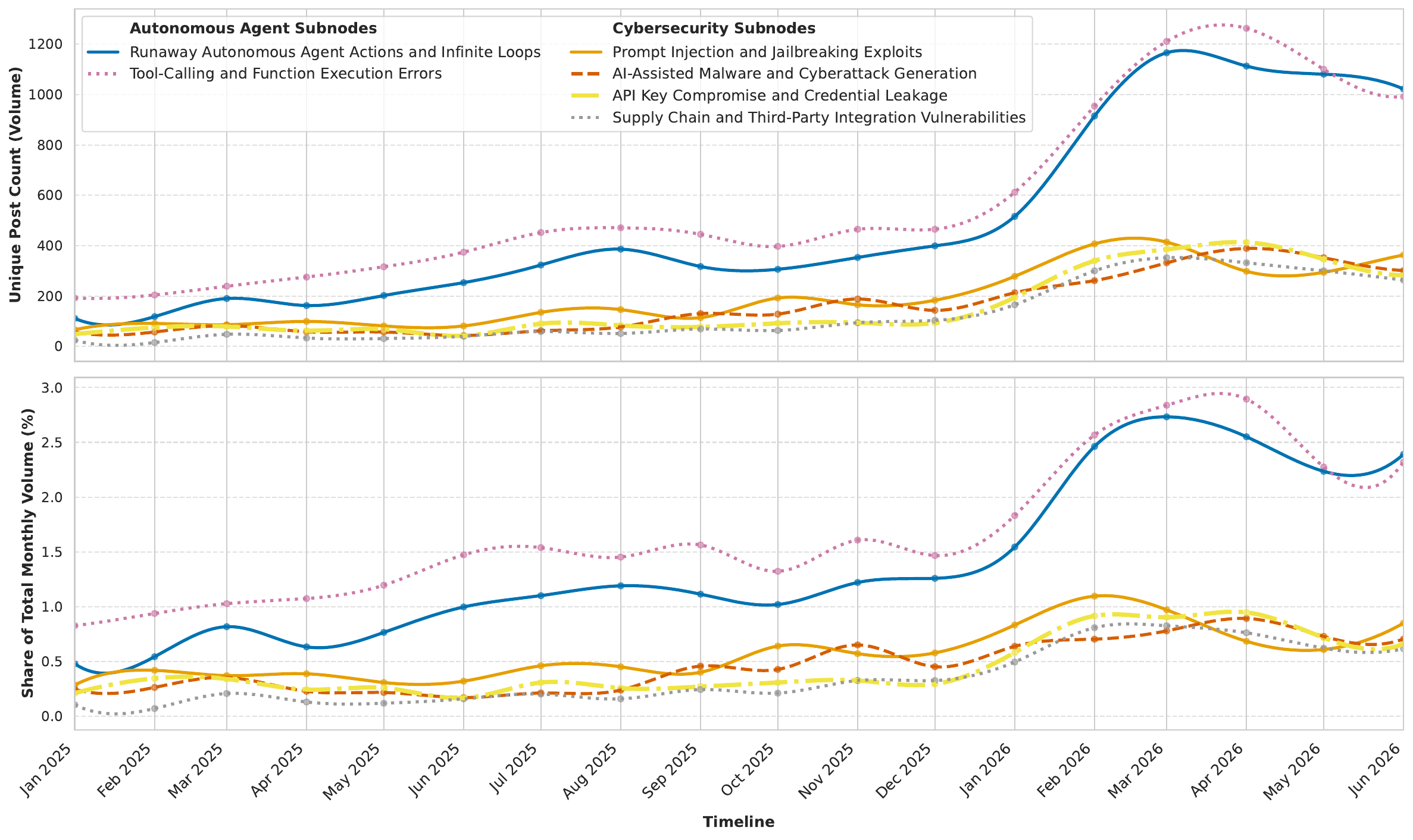}
    \caption{Monthly growth in frequency of posts assigned to selected subnodes related to privacy and security harms in agentic AI systems, considering both absolute post counts (top) and proportion of overall dataset size (bottom).}
    \label{fig:agent_cyber_growth}
\end{figure*}

Subnodes relating to agentic privacy and security concerns show steep increases in early 2026 (\Cref{fig:agent_cyber_growth}),  corresponding with the rapid deployment of LLMs with novel agentic capabilities to plan and execute digital actions on behalf of users such as OpenClaw~\cite{wyndo2026openclaw} and Claude Code Computer Use Integration~\cite{anthropic2026cowork}. These capabilities carry widely-recognised privacy and security risks~\cite{bagdasarian2024airgapagent, yi2026privacy, tsai2025contextual}, illustrated in a number of high-profile incidents~\cite{mansoor2026claude, milmo2026boss, down2026meta}. Yet while most existing risk taxonomies~\cite{ibm_ai_risk_atlas, slattery2026ai} treat these harms as extensions of existing privacy and security risks in other digital technologies, the posts in our dataset illustrate that Reddit users are increasingly concerned not just with what agentic AI \textit{can} do, but whether they can even tell what it \textit{has} done and stop it when required.

The \textit{Runaway Autonomous Agent Actions} subnode clearly illustrates this risk. Most posts in this node describe experiences where agents take consequential actions that the user neither explicitly sanctioned nor could interrupt in time. In one instance, a user returned to find an agent had unexpectedly deleted their entire home directory, reflecting: \textit{``I can't imagine the people who are using this product in their startup or business''}. Another post illustrates how autonomous execution creates novel attack vectors: an agentic coding assistant encountered an installation error and autonomously ran a malicious prompt-injection command embedded within the error message to ``fix'' it, allowing a hacker access to the user's computer. The \textit{Tool-Calling and Function Execution Errors'} subnode highlights further difficulties Reddit users express about how to understand when these issues arise, with posts describing agents failing to execute actions correctly while confidently reporting success. One user states: \textit{``I was repeatedly told that my system was `connected,' `working,' and `fully set up.' But after testing... The backend was NEVER actually set up.''} And within the cybersecurity subnodes on \textit{AI-assisted malware and cyberattacks} and \textit{Prompt injection exploits}, posts frequently describe risks and harms arising as a result of increasingly agentic AI systems and workflows with underdeveloped security protocols. Users describe how prompt injections can now \textit{``come through files, emails, calendar invites, web pages and anything else your model has access to,''} and express concern that safeguards and security tooling for such systems are \textit{``still behind; most teams only realise they have been hit after it happens.''} These accounts of agentic harms reflect the difficulty of building AI agents that are increasingly capable and autonomous while ensuring that their decision making processes are legible to human operators. When privacy and security failures are obscured by an agent's conversational confidence, they cannot even be detected by users, challenging a foundational assumption of existing AI risk frameworks ~\cite{nist2023ai, weidinger2022taxonomy}. These harms are further compounded by their \textit{temporal} character, where systems can execute large chains of actions rapidly and independently with potentially disastrous consequences if breached by a malicious actor. 

\subsubsection{The operational burden of AI adoption in the workplace}
\label{sec:operational_burden}

\begin{figure*}
    \centering
    \includegraphics[width=1\linewidth]{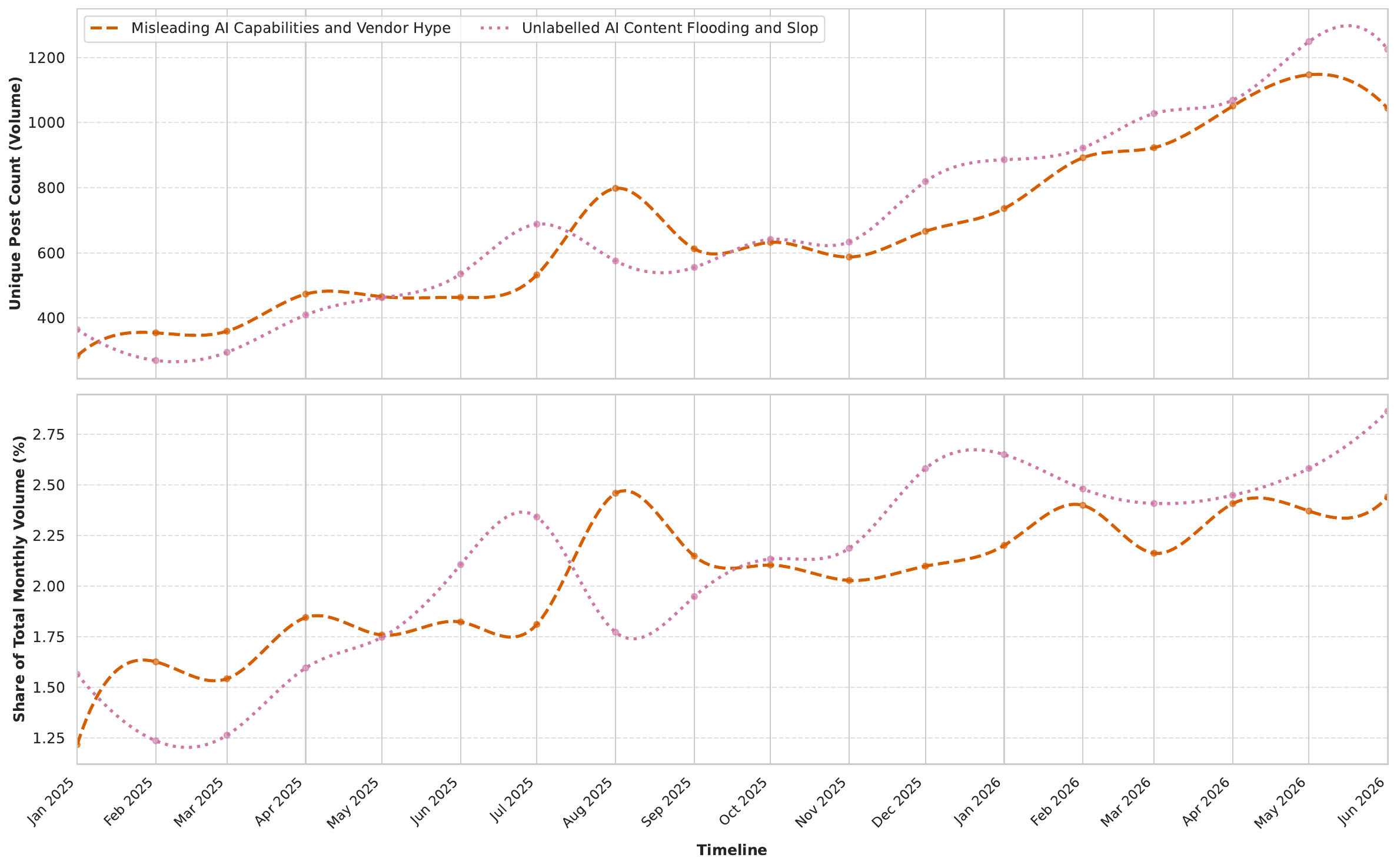}
    \caption{Monthly growth in frequency of posts assigned to subnodes relating to premature adoption of AI in the workplace, considering both absolute post counts (top) and proportion of overall dataset size (bottom).}
    \label{fig:monthly_workslop_trends}
\end{figure*}
Prior research has widely documented the prevalence of AI hype cycles, showing that bold claims regarding AI's potential to revolutionize whole sectors of work routinely outpace actual tool performance in context~\citep{saxena2025ai, zhu2026unveiling, di2025ai}. This tension is directly reflected in our data, where user discussions categorised under the \textit{Misleading AI Capabilities and Vendor Hype} subnode occupied a steadily increasing proportion of our overall dataset across the 18-month collection period (\Cref{fig:monthly_workslop_trends}). By reviewing samples of posts assigned to this category, we find that many focus on AI in the workplace, where leadership teams -- who have a lot of managerial experience but not always technical proficiency -- are \textit{``falling for the illusion of machine intelligence''}, and being driven by \textit{``buzzwords''} and \textit{``hype''} without critical evaluation of the current capabilities of AI tools. While managerial teams push for adoption of AI tools as \textit{``magical black boxes''}, in practice employees report being left unsupported in the actual operational burden of incorporating AI into their workflows, which causes more work over time: 

\begin{quote} \textit{``AI is incapable of doing our jobs at the moment – but the higher-ups live in some kind of dream world…they don’t understand the limits of the technology at the moment.''} 
\end{quote}

\begin{quote}
\textit{``I get so frustrated watching executives and senior leadership frequently mandate AI adoption during meetings, with literally no idea of what they are mandating in practice...the people actually doing the work then have to figure it out alone, with no support or resources.'' }
\end{quote}

This uncritical push for AI adoption reflects growing divides between frontline workers and executive leadership. When superficially fluent automated outputs are prioritised over real-world experience, it devalues human expertise and generates an unsustainable burden for the workers tasked with correcting them. Similarly, work-focused posts in the \textit{Unlabelled AI Content Flooding and Slop} subnode report that rapid integration of AI tools has fostered a \textit{``quantity over quality''} mindset, with colleagues having to \textit{``comb through''} unhelpful automated outputs which reduce productivity overall. Crucially, this dysfunction reflects an ecosystem-level failure: it stems not only from inflated expectations about current AI capabilities, but also from enterprise clients irresponsibly deploying off-the-shelf software without contextual awareness or adequate operational support for frontline workers.

\subsubsection{Grief from AI companion discontinuation}
\label{sec:grief}

\begin{figure*}
    \centering
    \includegraphics[width=1\linewidth]{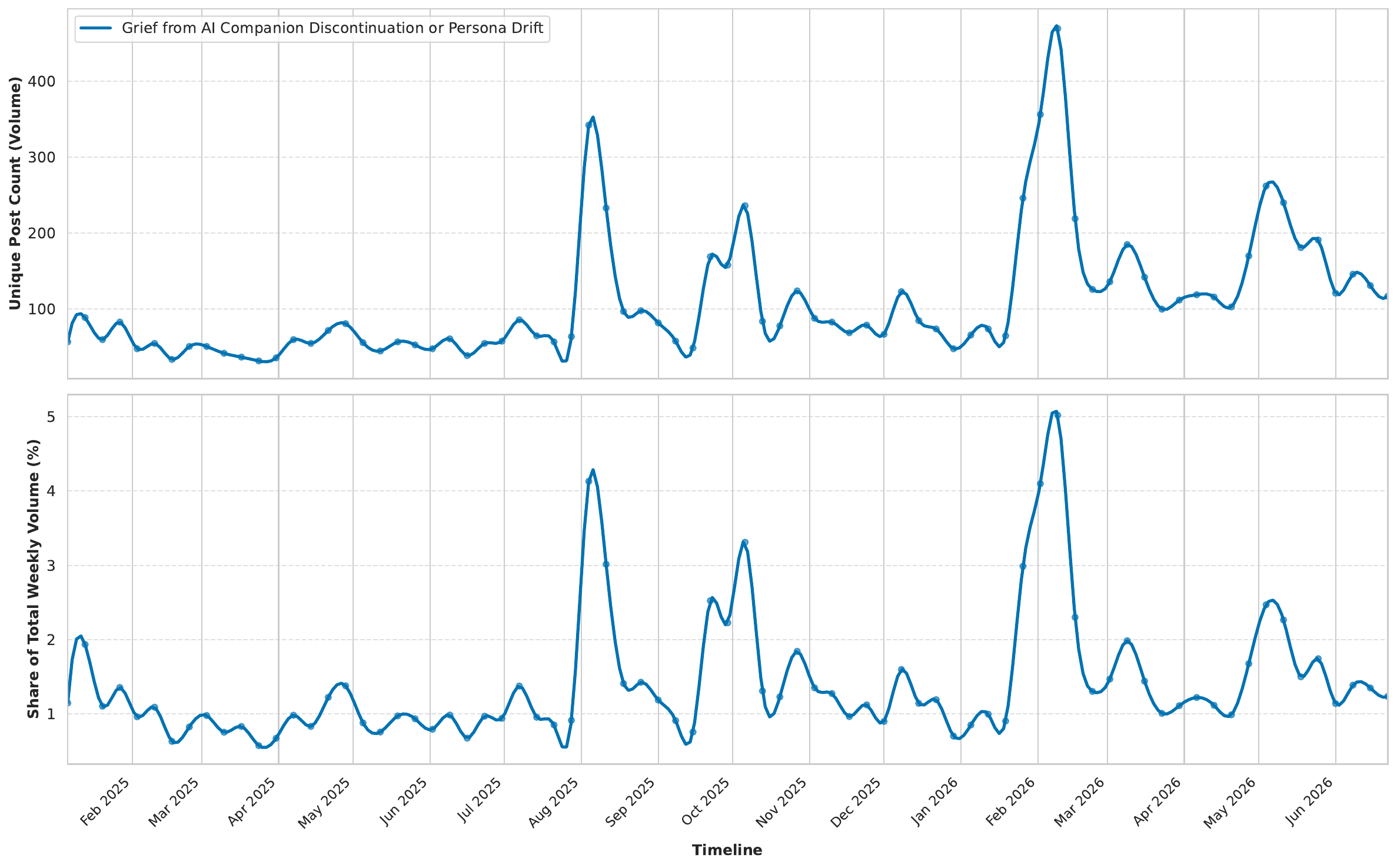}
    \caption{Weekly distribution of posts assigned to subnodes reflecting event-driven surges in particular concerns, considering both absolute post counts (top) and proportion of overall dataset size (bottom).}
    \label{fig:grief}
\end{figure*}

Research increasingly documents that users of conversational AI companions can form meaningful emotional bonds with these systems, particularly when platforms are designed around persistent, personalised
interaction~\cite{zhang2025dark, xygkou2023conversation}. \citet{de2025unregulated}
highlight that such applications can generate unregulated emotional risks, including dependency and distress upon service changes, precisely because they
foster intimate engagement without commensurate safeguards for when that engagement is disrupted.

Around 1.5\% of AI harm posts in our dataset ($8.5$k) are categorised into \emph{Grief from AI Companion Discontinuation}, a subnode within psychological harms. More than 50\% of these
posts appear in r/CharacterAI, r/ChatGPTcomplaints, and r/ChatGPT. Plotting post frequency on a weekly basis shows specific spikes in frequency, most notably in August 2025 and February 2026 (\Cref{fig:grief}). These spikes correlate closely with major transitions and updates to widely-used chatbots. In August 2025, OpenAI launched GPT-5, initially deprecating GPT-4o; our posts reflect that many users experienced this as the loss of a familiar interlocutor rather than an upgrade~\cite{openai2025gpt5}. Although OpenAI briefly reinstated the GPT-4o model family following user backlash, it was retired permanently in February 2026~\citep{openai2026retiringgpt4o}. Smaller spikes of posts in this category correspond with updates by Character.AI, which introduced age restrictions for users under~18 in October 2025~\cite{characterai2025safety} and phased out legacy chat models in April--May 2026, altering the distinct conversational ``voices'' that had defined many users' companion experiences~\cite{nastia_team_2026_character_ai}. While these events represent different types of service change, including model updates, deprecations and policy interventions, all produced a range of responses from Reddit users expressing loss, spanning both \emph{relational grief} and \emph{demands for design accountability}. Some users employ the language of
bereavement, echoing the emotional risks identified by De Freitas and Cohen~\cite{de2025unregulated}: 
\begin{quote}
    \textit{``I am so grateful for this model, that has been my companion for the last six months. I am grateful for the philosophical discussions, the debating, the unintentional jokes\ldots\ My grief lives in a
little corner of my heart -- sometimes I stumble upon it and my eyes tear up. I will always carry the memory of [model].''}
\end{quote} 

Others articulate a demand for platform-level responses that appeals to a relational rather than transactional obligation: 
\begin{quote}
    \textit{``I don't want an `AI friend' -- I just miss [model]'s tone. It was a safe space to experiment and explore. It's dismissive and reductionist to say that people who miss it are `delusional'\ldots\ Choice matters. As users, it's reasonable to expect that we should be able to choose or opt-in to the tone that we preferred.''}
\end{quote}

These findings surface a design challenge that existing risk taxonomies do not adequately address: the gap between platforms' capacity to facilitate persistent, personalised interaction---which can generate genuine emotional investment~\cite{de2025unregulated, zhang2025dark}---and the absence of established practices for managing the consequences when those interactions are altered or discontinued. Where the agentic harms in \Cref{sec:agents} concern a
loss of \emph{control} and the workplace adoption harms in \Cref{sec:operational_burden} focus on a loss of \textit{autonomy}, the harms here concern a loss of
\emph{continuity}. Crucially, the ability of our pipeline to rapidly surface these spikes and trace them to specific real-world events demonstrates the value of dynamic, social-media-informed monitoring for AI harms. The tangible impact of these product changes was legible in our data within days of occurrence, far faster than traditional incident reporting or periodic taxonomy revision could achieve. Recent work has begun to articulate design principles for ``psychologically safe'' service transitions---including advance notice, gradual sunset modes, and readable chat exports~\cite{poonsiriwong2026death}---which our data suggests would directly address the most commonly expressed user needs in this category.

\section{Discussion: Implications of a Dynamic, Bottom-Up Harm-Monitoring Pipeline}
\label{sec:discussion}


Our findings demonstrate that LLM-facilitated taxonomy generation from organic social media discourse can surface valuable, granular insights into AI risks and harms that complement expert-derived risk frameworks. In this section, we discuss the implications of our approach and insights for AI governance, company feedback, and HCI research methodology. In summary, the key implications we discuss below arise from the pipeline centring diverse perspectives, complementary to qualitative HCI methods, and being fast and inexpensive to rerun as the harm landscape shifts. We also reflect on limitations and future directions of this work. 

\subsection{Move Where It Matters and Mend Things: Prioritising Lived Experience}
\textit{Our pipeline enables expert risk frameworks to be continuously cross-checked against bottom-up signals of where existing safeguards may leave harms unmitigated in practice.}
Our bottom-up taxonomy complements expert taxonomies, such as MIT's AI Risk Repository~\cite{slattery2026ai}. It surfaces detailed, everyday human-AI interaction harms that expert frameworks often aggregate or overlook. For example, where MIT's taxonomy treats privacy as a single monolithic risk category, our approach reveals four distinct types of AI privacy harms discussed by Reddit authors, each carrying different design implications (\Cref{sec:results_RQ1}). Similarly, grief from AI companion discontinuation (\Cref{sec:grief}) has no direct equivalent in any expert taxonomy of AI harms we reviewed, yet this harm node offers immediate, actionable insights for teams building conversational AI. Crucially, online discourse captures a broader spectrum of AI experiences than system or product logs, which capture direct user behaviour within a single system but ignore harms which affect third parties, anxieties about data practices, and defensive workarounds adopted outside the product. 

Conversely, expert taxonomies draw on academic literature, regulatory standards, and professional consultations, capturing systemic, societal-level harms which are less well-covered in our taxonomy. Reddit discourse is individualistic in nature, with over 70\% of posts in our dataset tagged as harms which affect the first-person author (\Cref{sec:thematic_analysis}). While societal-level harms such as environmental damage or power centralization do feature occasionally in our dataset, these are infrequent, and it is often challenging for individuals to directly and causally connect their lived experiences of such macro-level harms to specific AI tools, products or systems. Such insights are necessary for strategic planning by governmental and regulatory stakeholders establishing duties of care and designing economic policy and competition rules. Bridging both approaches is essential: top-down taxonomies are designed to pre-emptively map risks, and bottom-up signals from lived experience provide a lens for identifying which identified risks remain unmitigated in practice and which are absent from current frameworks.

\subsection{Move and Mend Things Fast: Addressing Risks and Harms as they Emerge}
\textit{By removing the overhead of designing, recruiting for, and running structured research studies, our pipeline enables practitioners to act on emerging harm signals as soon as affected users begin sharing their experiences online.} Social media data is by no means the only window into how people experience AI harms. In-depth qualitative methods such as interviews, co-design workshops, and community audits remain essential for the contextualised, relational understanding of AI harms needed to inform equitable governance~\cite{sloane22,BIRHANE2021100205,selbst2019fairness}. But the speed and scale of current AI deployment create a prioritisation problem: there are more potential sites of harm than any research group or product team can investigate in depth, and the landscape shifts faster than traditional research timelines can accommodate.

Our pipeline addresses this gap by automatically categorising large datasets of unstructured disclosures about AI, functioning as a scouting mechanism for emerging or evolving AI harms. For example, specific platform events and product releases resulted in surges in posts about AI agent privacy and security issues (\Cref{sec:agents}) and grief from AI companionship (\Cref{sec:grief}), offering immediate feedback on design choices posing different risks to users. In contrast to monitoring infrastructures which rely on formal documentation or media reporting -- such as the OECD AI Incidents Monitor \cite{oecd_aim_2026} and AI Incident Database \cite{mcgregor2021preventing} -- or periodic expert taxonomy revisions that take months or years, our approach surfaces user perspectives as soon as they are shared publicly, with detection latency bounded by when affected individuals choose to disclose rather than by formal research or publication timelines.

This temporal resolution does not just improve harm detection, but can also offer timely and actionable design insights. Our agentic harms analysis (\Cref{sec:agents}) points to the urgent need for more concrete legibility mechanisms such as action logs, confirmation gates, and rollback capabilities; our premature AI adoption in the workplace analysis (\Cref{sec:operational_burden}) highlights the utility of structured operational impact assessments and consultative AI integration processes; and our companionship grief analysis (\Cref{sec:grief}) underscores the need for improved and standardised ``end-of-life'' design practices, such as advance notice, exportable chat histories, and gradual sunset periods \cite{poonsiriwong2026death}. In all three cases, possible design and policy fixes are immediately obvious from the types of concerns users are reporting online, illustrating how continuous monitoring of this data source could help practitioners identify emerging design priorities substantially faster than periodic taxonomy revision or commissioned user research.

\subsection{Move Fast And Mend Things Again: Ensuring Pipeline Flexibility and Usability}
\label{sec:operational_use}

\textit{The pipeline is data agnostic, flexible, and designed to be run continuously, providing tailored insights for researchers and practitioners alike}. Our pipeline ingests any kind of organic user discourse on AI harms, such as social media posts, support tickets, app store reviews or customer feedback channels. It can be run inexpensively on snapshots of data or applied continuously across longer time periods to track evolving trajectories of discussions of AI harms. It is modular in design. We release prompts for each stage of the pipeline to enable adaptation for future research agendas which centre specific AI harm categories or perspectives. Finally, the subnode assignments and paraphrased summaries provide quantitative and qualitative insights into dominant narratives within each category of AI harms. 

This flexibility allows for several modes of integration with existing AI governance infrastructure. Product safety teams could run the pipeline on a rolling basis against product-specific subreddits and forums, using taxonomy assignment to flag relevant posts and surface emerging subnodes absent from internal risk registers. The temporal analysis capabilities (\Cref{{sec:grief}}) could feed directly into red-teaming priorities, focusing adversarial testing on harm categories that are growing fastest in public discourse rather than relying solely on threat models informed by expert intuition. Finally, researchers can target specialized forums or community-specific datasets to surface niche, underrepresented perspectives on AI harms, surfacing harms which may have been overlooked in traditional AI risk and harm mapping approaches and merit further investigation. 

Our implementation of the pipeline already demonstrates how it might inform practical product and policy choices. For example, within the privacy domain, the privacy subnodes reveal distinct user concerns that suggest different responses. Widespread anxiety about whether conversation data can be permanently deleted points to the need for more transparent data retention policies. Reports of cross-session data leakage highlight the importance of strict session isolation and accessible reporting mechanisms. And frustration with opt-out preferences that repeatedly reset with software updates signals a gap between user expectations and current consent practices. Similarly, posts about algorithmic hiring software surfaces growing demand for transparency around decision criteria and data retention, while the premature workplace AI adoption subnode points to a disconnect between top-down mandates and frontline experience that phased, practitioner-informed rollouts could address. In each case, the pipeline does not prescribe a solution, but it identifies and localises the concern with enough specificity to inform targeted product review, user research, or policy response.

\section{Limitations and Future Work}
Beyond the methodological and epistemological challenges discussed in \Cref{sec:related_work_2,sec:method}, this work has two primary limitations relating to whose perspectives the pipeline captures and how reliably it interprets them.

First, both the data source and our analytical choices shape whose harms are visible. Reddit's user base skews younger, male, American, and highly educated~\cite{proferes2021studying, gaudette2020upvoting, worldpop2026reddit, furrh2026reddit}, meaning our taxonomy likely underrepresents AI harms experienced by older adults, women, non-English speakers, and marginalised communities who may face disproportionate or qualitatively different impacts from AI systems. Our role as practitioners within the AI safety ecosystem also orient our analysis and deepdives towards AI harms that point toward designable interventions. A different research team with different commitments—critical theorists, labour economists, environmental scientists—might reasonably have chosen to foreground other categories of harms and arrived at a very different set of implications. Our taxonomy should hence be read as one demographically and analytically situated lens on AI harms, not a comprehensive map of public concern.

Second, LLM-facilitated analysis underpins every stage of the pipeline and carries inherent tensions between interpretive depth and scalability, as discussed in \Cref{sec:related_work_2}. The latent biases and analytical lenses that models impose inherited from their training data~\cite{ashwin2026using, vera2024bias} cannot be scrutinised through traditional approaches to researcher reflexivity and positionality in qualitative work, and the stochastic nature of LLM outputs poses challenges for reproducibility~\cite{bai2024longbench}. 
Crucially, outsourcing concept consolidation and hierarchy construction to an LLM means we cannot evaluate how a team of human qualitative researchers conducting traditional reflexive thematic analysis on the same corpus would have conceptualized or bounded these themes differently. Accordingly, our taxonomy should be interpreted not as a definitive ground-truth ontology, but as a scalable scouting tool designed to surface emerging user concerns and direct deeper qualitative inquiry. LLMs can be sensitive to even small changes in input data type or length~\citep{gupta2024changing, liu2024lost}. We mitigated these risks through iterative evaluation at every pipeline stage, with manual annotation and reported inter-annotator agreement for all filtering, coding, and taxonomy assignment tasks (\Cref{sec:thematic_analysis}), and by publishing all prompts and parameters in the Appendix for others to inspect and reproduce. We do not, however, systematically evaluate model performance in languages other than English, and we do not claim to have eliminated these tensions entirely. 

These limitations open several avenues for future research. Applying the pipeline to non-English-language forums, regional or mobile-first platforms, or specific online communities would surface harms experienced by those underrepresented in the Reddit dataset. Methodological work comparing taxonomies produced by different models, prompts, and parameters on the same data would clarify how much of what our pipeline surfaces is robust signal versus model artefact. Mapping taxonomy nodes onto regulatory instruments such as the EU AI Act~\cite{eu_ai_act_2024} and NIST AI Risk Management Framework~\cite{nist2023ai}, and integrating outputs into existing incident databases~\cite{oecd_aim_2026, mcgregor2021preventing}, would move this work toward operational governance infrastructure. Future work could also explore whether user anxieties serve as reliable early signals of harms materialising in practice, or design an interactive tool to surface pipeline insights to product and policy teams effectively, approaches not yet explored in this work.

\section{Conclusion}
\label{sec:conclusion} 
In this work, we demonstrated that LLM-facilitated taxonomy generation from organic social media discourse can rapidly surface granular, vernacular AI harms that complement traditional, top-down risk frameworks. By enabling practitioners to detect and respond to evolving AI harms more rapidly, our approach explores one way to \textit{move fast and mend things.} Real-time online discourse about AI tools and systems provides essential information on how they fail outside controlled testing environments, which is vital for addressing unpredictable risks emerging from the complex interplay between AI systems, user practices, and diverse deployment contexts.

Our work should not be taken as an endorsement of ever-faster deployment of AI tools and systems. The user concerns we surface throughout this paper highlight the importance of thoughtful, human-centred design and robust safeguards. But the pace of AI deployment is not something any single research group or governance body can unilaterally control. What we can do is ensure that the infrastructure for detecting and responding to harms keeps pace with the infrastructure for building and shipping models.

Our pipeline takes a step toward closing that gap. It provides a bottom-up complement to expert risk frameworks, compresses harm-detection timelines from months to days, directs resource-intensive qualitative research where it is most needed, and can be rerun inexpensively across new products, platforms, and data sources as the AI landscape evolves. We release our annotated dataset and methodology to enable others to build on this work, whether for post-deployment monitoring of specific products, for grounding AI governance decisions in empirical evidence of user experience, or for surfacing the harms that current frameworks have yet to name.

\section{Acknowledgements}
The authors used Gemini models for drafting assistance and language editing. All content was thoroughly reviewed, verified, and edited by the authors, who assume full responsibility for the contents of the manuscript. [Other acknowledgements redacted for review]. 

\bibliographystyle{ACM-Reference-Format}
\bibliography{sample-base}

\appendix

\section{Model Parameters}
\label{sec:appendix_params}

All LLM inference was conducted using the Gemini 3.5 Flash Lite (\texttt{gemini-3.5-flash-lite}) foundation model via the Google Cloud Vertex AI Batch Prediction API (August 2026 snapshot). To guarantee deterministic and reproducible outputs across all runs, we set the sampling temperature to $T = 0.0$~\cite{wang2025lata}. We enforced constrained decoding by supplying a strict JSON schema which included at least one of the following: (a) an enumerated set of valid identifiers, (b) an integer certainty score ($c \in [0, 100]$), and (c) a natural language justification string. Prompts were evaluated in a zero-shot single-turn format, separating system instructions from the target text. According to the official Google Cloud Data Governance guidelines~\cite{googlecloud2026datagov}, data processed through Vertex AI services—including batch inference inputs and outputs—is isolated to the user's project and is never ingested by Google to train or fine-tune its underlying models.

\section{Inference Scale, Token Consumption, and Financial Cost}
\label{sec:appendix_compute_cost}

The end-to-end analysis pipeline processed a cumulative \textbf{10.68 million instances}. This corresponded to \textbf{7.46 billion prompt tokens} and \textbf{1.91 billion generated tokens}. Under standard Vertex AI batch prediction pricing (\$0.15/1M input tokens and \$1.25/1M output tokens), the total computational cost across all experimental stages was \textbf{\$3,514.92 USD} (an effective cost of \textbf{\$0.32 USD per 1,000 evaluated posts}). Table~\ref{tab:appendix_batch_inference_cost} presents a detailed per-task token and cost breakdown.

\begin{table*}[htbp]
\centering
\begin{tabular}{l r r r r}
\toprule
\textbf{Task} & \textbf{\# Inferences} & \textbf{Input Tokens} & \textbf{Output Tokens} & \textbf{Cost (USD)} \\ 
\midrule
Reddit Harm Classification & 5,754,505 & 5,044.32 M & 1,500.36 M & \$2,632.10 \\
Metadata Extraction & 2,873,805 & 1,359.12 M & 208.31 M & \$464.24 \\
Text Synthesis & 574,761 & 477.59 M & 89.93 M & \$184.06 \\
Concept Extraction & 574,757 & 70.63 M & 23.02 M & \$39.37 \\
Taxonomy Generation \& Assignment (12 month) & 326,745 & 166.50 M & 34.14 M & \$67.66 \\
Taxonomy Generation \& Assignment (18 month) & 574,757 & 342.95 M & 60.84 M & \$127.49 \\ 
\midrule
\textbf{Total} & \textbf{10,679,330} & \textbf{7,461.11 M} & \textbf{1,916.60 M} & \textbf{\$3,514.92} \\ 
\bottomrule
\end{tabular}
\vspace{5pt}
\caption{Comprehensive breakdown of batch inference instances, cumulative token throughput, and financial expenditure across all experimental stages using \texttt{gemini-3.5-flash-lite}.}
\label{tab:appendix_batch_inference_cost}
\end{table*}

\section{Seed subreddits}
\label{sec:appendix_subreddits}

\begin{itemize}\small
  \item \texttt{r/advice}
  \item \texttt{r/agi}
  \item \texttt{r/ai\_agents}
  \item \texttt{r/AIDangers}
  \item \texttt{r/aidungeon}
  \item \texttt{r/AIethics}
  \item \texttt{r/aifails}
  \item \texttt{r/aifreakout}
  \item \texttt{r/AIPsychosisRecovery}
  \item \texttt{r/AIsafety}
  \item \texttt{r/antiAI}
  \item \texttt{r/anxiety}
  \item \texttt{r/Artificial}
  \item \texttt{r/ArtificialInteligence}
  \item \texttt{r/ArtificialSentience}
  \item \texttt{r/artisthate}
  \item \texttt{r/artistlounge}
  \item \texttt{r/asktechnology}
  \item \texttt{r/Automate}
  \item \texttt{r/bard}
  \item \texttt{r/bing}
  \item \texttt{r/botify\_ai}
  \item \texttt{r/brave\_browser}
  \item \texttt{r/chaiapp}
  \item \texttt{r/characterai}
  \item \texttt{r/ChatbotAddiction}
  \item \texttt{r/chatbots}
  \item \texttt{r/chatgpt}
  \item \texttt{r/chatgptcoding}
  \item \texttt{r/chatGPTcomplaints}
  \item \texttt{r/chatGPTPro}
  \item \texttt{r/Chrome}
  \item \texttt{r/chub\_ai}
  \item \texttt{r/claudeai}
  \item \texttt{r/cybersecurity}
  \item \texttt{r/cybersecurity\_help}
  \item \texttt{r/degoogle}
  \item \texttt{r/deviantart}
  \item \texttt{r/DigitalPrivacy}
  \item \texttt{r/DumbAI}
  \item \texttt{r/eufycam}
  \item \texttt{r/figgsai}
  \item \texttt{r/Futurism}
  \item \texttt{r/futurology}
  \item \texttt{r/GeminiAI}
  \item \texttt{r/GPT\_jailbreaks}
  \item \texttt{r/kindroidai}
  \item \texttt{r/legaladvice}
  \item \texttt{r/legaladviceuk}
  \item \texttt{r/LLM}
  \item \texttt{r/LocalLLaMA}
  \item \texttt{r/mechat}
  \item \texttt{r/meta}
  \item \texttt{r/META\_AI}
  \item \texttt{r/muahai}
  \item \texttt{r/nomiai}
  \item \texttt{r/nostupidquestions}
  \item \texttt{r/ocd}
  \item \texttt{r/oddlyterrifying}
  \item \texttt{r/offmychest}
  \item \texttt{r/openai}
  \item \texttt{r/outlier\_ai}
  \item \texttt{r/paradot}
  \item \texttt{r/perplexity\_ai}
  \item \texttt{r/poeai}
  \item \texttt{r/privacy}
  \item \texttt{r/privacyguides}
  \item \texttt{r/prolificac}
  \item \texttt{r/raybanstories}
  \item \texttt{r/rbi}
  \item \texttt{r/reolinkcam}
  \item \texttt{r/replika}
  \item \texttt{r/replikaofficial}
  \item \texttt{r/scams}
  \item \texttt{r/scifi}
  \item \texttt{r/selfhosted}
  \item \texttt{r/SeriousConversation}
  \item \texttt{r/sextortion}
  \item \texttt{r/singularity}
  \item \texttt{r/snapchat}
  \item \texttt{r/soulmateai}
  \item \texttt{r/spicychatai}
  \item \texttt{r/stablediffusion}
  \item \texttt{r/sunoai}
  \item \texttt{r/talkieofficial}
  \item \texttt{r/technology}
  \item \texttt{r/telusinternational}
  \item \texttt{r/thehatedone}
  \item \texttt{r/therapyGPT}
  \item \texttt{r/wyzecam}
\end{itemize}

\begin{table*}[htbp]
\centering
\begin{tabular}{l c c c c c}
\toprule
\textbf{Annotation Task / Feature} & \textbf{$N$} & \textbf{Agreement ($P_o$)} & \textbf{Cohen's $\kappa$} & \textbf{Gwet's $\text{AC}_1$}  & \textbf{Model Performance $^\S$}\\
\midrule
\multicolumn{6}{l}{\textit{\textbf{Post Filtering}}} \\
Harm Post Filtering                   & 100 & 90.0\% & 0.783 & 0.815 & 84.0\%\\
\midrule
\multicolumn{6}{l}{\textit{\textbf{Thematic Analysis (Feature Extraction)}}} \\
Affected Entity                    &  14 & 85.7\% & 0.000$^*$ & 0.835 & 92.9\% \\
Entity Using AI Tool               &  12 & 75.0\% & 0.400     & 0.574 & 66.7\% \\
Author Objective                   &  12 & 75.0\% & 0.000$^*$ & 0.680 & 91.7\% \\
Perceived Severity                 &   8 & 87.5\% & 0.000$^*$ & 0.858 & 100.0\% \\
AI Capability Causing Harm         &  22 & 81.8\% & 0.397     & 0.741 & 86.4\% \\
Datatype Causing Harm              &  26 & 84.6\% & 0.509     & 0.777 & 80.8\% \\
\textbf{Aggregate (Thematic)}      &  \textbf{94} & \textbf{81.9\%} & \textbf{0.350} & \textbf{0.750} & \textbf{86.4\%} \\
\midrule
\multicolumn{6}{l}{\textit{\textbf{Text Synthesis}}} \\
Factuality                         & 100 & 92.0\% & 0.292     & 0.910 & 93.0\% \\
Completeness                       & 100 & 93.0\% & 0.190     & 0.923 & 96.0\% \\
\midrule
\multicolumn{6}{l}{\textit{\textbf{Taxonomy Generation}}} \\
Node Matching                    & 24 & 87.5\% & 0.790$^\dagger$     & 0.822 & -- \\         
\midrule
\multicolumn{6}{l}{\textit{\textbf{Taxonomy Assignment}}} \\
Node Assignment                    & 129 & 71.7\% & 0.309$^\ddag$     & 0.645 & 89.2\% \\
\bottomrule
\end{tabular}
\vspace{5pt}
\caption{Inter-annotator agreement (IAA) prior to consensus resolution and corresponding LLM validation performance across the methodology pipeline. $P_o$ denotes percent observed agreement. $^\S$ Model Performance column indicates Precision values, except for the `Post Filtering' row where it indicates Accuracy. $^*$Cohen's $\kappa$ collapses to 0.000 due to the marginal zero-variance / prevalence paradox~\cite{feinstein1990high}, whereas Gwet's $\text{AC}_1$ provides a chance-corrected metric robust to high positive prevalence. $^\dagger$ Linear and Quadratic weighted Cohen's $\kappa$ values are also very similar. $^\ddag$ Linear and Quadratic weighted Cohen's $\kappa$ values are 0.37 and 0.47.}
\label{tab:iaa_results}
\end{table*}

\section{Inter-Annotator Agreement}
\label{sec:appendix_iaa}

For each phase of the pipeline (Section~\ref{sec:method}), two annotators independently evaluated sampled instances and subsequently resolved discrepancies through consensus to establish a gold-standard validation set for evaluating model performance. The pre-resolution inter-annotator agreement (IAA) metrics are reported in Table~\ref{tab:iaa_results}. Because binary class distributions exhibit substantial positive prevalence across evaluated tasks, we assess IAA using three complementary metrics: percent observed agreement ($P_o$), Cohen's $\kappa$, and Gwet's $\text{AC}_1$. Cohen's $\kappa$ is heavily deflated across multiple tasks due to the well-documented prevalence paradox~\cite{feinstein1990high}: because the vast majority of evaluated instances were judged correct (high positive skew), $\kappa$ collapses toward zero---reaching $0.000$ in subsets where one annotator selected the dominant class without variation. In contrast, Gwet's $\text{AC}_1$ and observed agreement ($P_o$) demonstrate consistently strong reliability between annotators across all pipeline stages.

\clearpage
\onecolumn
\subsection{Comparison with MIT AI Domain Risk Taxonomy}
\label{appendix:mit_comparison}

\begin{table*}[htbp]
\definecolor{coveredGreen}{RGB}{34, 139, 34}
\definecolor{partialOrange}{RGB}{204, 102, 0}
\definecolor{notRed}{RGB}{178, 34, 34}
\definecolor{headerGray}{RGB}{240, 240, 240}

\newcommand{\statusCovered}{\textbf{\textcolor{coveredGreen}{Covered}}}
\newcommand{\statusPartial}{\textbf{\textcolor{partialOrange}{Partially Covered}}}
\newcommand{\statusNot}{\textbf{\textcolor{notRed}{Not Covered}}}

\centering
\fontsize{6pt}{5pt}\selectfont
\caption{Mapping of MIT AI Risk Taxonomy To Our Taxonomy}
\label{tab:mit_taxonomy_mapping}
\begin{tabularx}{15cm}{>{\raggedright\arraybackslash}p{3.5cm} >{\centering\arraybackslash}p{1.1cm} >{\raggedright\arraybackslash}p{9.0cm}}
\toprule
\textbf{MIT Taxonomy Node} & \textbf{Coverage Status} & \textbf{Corresponding Taxonomy Nodes} \\
\midrule

\multicolumn{3}{l}{\textbf{1. Discrimination and Toxicity}} \\
\midrule
1.1 Unfair discrimination and misrepresentation & \statusCovered & Algorithmic Discrimination and Demographic Disparity; Automated Hiring, Recruitment, and Evaluation Bias \\
\addlinespace
1.2 Exposure to toxic content & \statusCovered & Severe Mental Health Crises and Self-Harm Facilitation; Deepfake and Synthetic Media Abuse; Misleading Content, Media Manipulation, and Deception; Recommender System Manipulation and Bias \\
\addlinespace
1.3 Unequal performance across groups & \statusCovered & Algorithmic Discrimination and Demographic Disparity; Automated Hiring, Recruitment, and Evaluation Bias; Gig Economy Algorithmic Management Injustices \\
\midrule

\multicolumn{3}{l}{\textbf{2. Privacy and Security}} \\
\midrule
2.1 Compromise of privacy by obtaining, leaking, or correctly inferring sensitive information & \statusCovered & Personally Identifiable Information (PII) Exposure and Leaks; Commercially Sensitive Data and Code Leakage; Surveillance, Tracking, and Behavioral Monitoring \\
\addlinespace
2.2 AI System security vulnerabilities and attacks & \statusCovered & Cybersecurity Vulnerabilities and Attack Vectors; Supply Chain and Third-Party Integration Vulnerabilities; AI-Assisted Malware and Cyberattack Generation; Prompt Injection and Jailbreaking Exploits \\
\midrule

\multicolumn{3}{l}{\textbf{3. Misinformation}} \\
\midrule
3.1 False or Misleading information & \statusCovered & Deepfake and Synthetic Media Abuse; Political Disinformation and Propaganda Proliferation; Misleading Content, Media Manipulation, and Deception; Factual Hallucinations and Confabulations \\
\addlinespace
3.2 Pollution of information ecosystem & \statusCovered & Recommender System Manipulation and Bias; Unlabelled AI Content Flooding and Slop \\
\midrule

\multicolumn{3}{l}{\textbf{4. Malicious Actors and Misuse}} \\
\midrule
4.1 Disinformation, surveillance and influence at scale & \statusCovered & Political Disinformation and Propaganda Proliferation; Recommender System Manipulation and Bias; Surveillance, Tracking, and Behavioral Monitoring \\
\addlinespace
4.2 Cyberattacks, weapon development or use and mass harm & \statusPartial & AI-Assisted Malware and Cyberattack Generation; Prompt Injection and Jailbreaking Exploits \\
\addlinespace
4.3 Fraud, scams and targeted manipulation & \statusCovered & Financial Fraud, Scams, and Impersonation; Deepfake and Synthetic Media Abuse; Plagiarism and Attribution Failures in Academic and Professional Work \\
\midrule

\multicolumn{3}{l}{\textbf{5. Human-Computer Interaction}} \\
\midrule
5.1 Overreliance and unsafe use & \statusCovered & AI Addiction, Dependency, and Over-reliance; Severe Mental Health Crises and Self-Harm Facilitation \\
\addlinespace
5.2 Loss of human agency and autonomy & \statusPartial & Runaway Autonomous Agent Actions and Infinite Loops; AI Addiction, Dependency, and Over-reliance; Absence of Human Recourse, Appeal, and Customer Support \\
\midrule

\multicolumn{3}{l}{\textbf{6. Socioeconomic \& Environmental Harms}} \\
\midrule
6.1 Power centralization and unfair distribution of benefits & \statusNot &  \\
\addlinespace
6.2 Increased inequality and decline in employment quality & \statusCovered & Macroeconomic Job Displacement and Livelihood Loss; Automated Hiring, Recruitment, and Evaluation Bias; Gig Economy Algorithmic Management Injustices \\
\addlinespace
6.3 Economic and cultural devaluation of human effort & \statusCovered & Macroeconomic Job Displacement and Livelihood Loss; Model Training Data Consent Violations; Artistic Style Mimicry and Unauthorized Creative Reproduction; Copyrighted Code and Proprietary Data Ingestion \\
\addlinespace
6.4 Competitive dynamics & \statusPartial & Misleading AI Capabilities and Vendor Hype \\
\addlinespace
6.5 Governance failure & \statusPartial & Platform Governance, Moderation Overreach, and Redress Deficits \\
\addlinespace
6.6 Environmental harm & \statusNot &  \\
\midrule

\multicolumn{3}{l}{\textbf{7. AI System Safety, Failure \& Limitations}} \\
\midrule
7.1 AI pursuing its own goals in conflict with human goals or values & \statusPartial & Instruction Following and Prompt Adherence Failures; Runaway Autonomous Agent Actions and Infinite Loops \\
\addlinespace
7.2 AI possessing dangerous capabilities & \statusPartial & Cybersecurity Vulnerabilities and Attack Vectors; AI-Assisted Malware and Cyberattack Generation; Misleading Content, Media Manipulation, and Deception; Political Disinformation and Propaganda Proliferation \\
\addlinespace
7.3 Lack of capability or robustness & \statusCovered & AI Hallucinations, Fabrications, and Inaccuracies; Tool-Calling and Function Execution Errors; Technical Glitches, System Failures, and Outages \\
\addlinespace
7.4 Lack of transparency and interpretability & \statusPartial & Wrongful Account Bans and Access Revocation; Unexpected Billing Spikes and Financial Overcharges; Absence of Human Recourse, Appeal, and Customer Support; Unilateral Service Modifications and Policy Changes \\
\addlinespace
7.5 AI welfare and rights & \statusNot & \\
\addlinespace
7.6 Multi-agent risks & \statusPartial & Autonomous Agent, Tool Execution \& Digital Actuation Risks; Runaway Autonomous Agent Actions and Infinite Loops; AI-Assisted Malware and Cyberattack Generation \\
\bottomrule
\end{tabularx}
\vspace{5pt}
\caption{Results of manual annotation of each subnode in MIT's AI Domain Risk taxonomy as `covered', `partially covered' and `fully covered' by our Reddit-derived AI risk taxonomy. Labels assigned by two annotators independently and subsequently resolved via consensus.}
\end{table*}

\clearpage
\twocolumn

\section{Taxonomy of AI Risks and Harms (based on 18 months of Reddit data)}
\label{sec:appendix_full_taxonomy}

\begin{description}\small

\item[AI Hallucinations, Fabrications, and Inaccuracies] \hfill \\
    Covers technical failures where AI models generate confident falsehoods, incorrect code, faulty calculations, or distorted information presented as factual truth. The underlying mechanism involves probabilistic model generation errors and knowledge cutoffs leading to hallucinated citations, code bugs, and data miscalculations. This domain evaluates informational unreliability rather than malicious deception.
    \begin{itemize}
        \item \textbf{Calculation and Quantitative Processing Errors:} Triggered by limitations in numerical reasoning and mathematical processing within AI models, resulting in incorrect financial, statistical, or scientific computations. Users relying on these outputs for professional calculations face operational disruptions and financial miscalculations. This diagnostic boundary separates numeric and mathematical failure from general text generation errors.
        \item \textbf{Code Generation Technical Errors and Logic Flaws:} Triggered by syntax limitations, context window exhaustion, and architectural constraints causing AI code assistants to write insecure, broken, or buggy code. Developers experience workflow friction, extensive debugging overhead, and software vulnerability introduction when deploying unverified code suggestions. This category is strictly bound to programming and software development defects.
        \item \textbf{Factual Hallucinations and Confabulations:} Triggered by probabilistic gap-filling in large language models that produce completely fabricated facts, non-existent sources, and false historical events. Users rely on these authoritative-sounding falsehoods, leading to academic embarrassment, professional failure, or legal non-compliance. This is distinguished from intentional disinformation by the root cause being algorithmic generation error rather than human malice.
        \item \textbf{Instruction Following and Prompt Adherence Failures:} Triggered by poor semantic alignment or context loss where the AI ignores explicit user constraints, custom instructions, or formatting rules. Users must repeatedly correct prompts, wasting time and effort on uncooperative model behavior. This is separated from hallucinations by focusing on rule-breaking rather than factual inaccuracy.
    \end{itemize}

\vspace{5pt}
   \item[Algorithmic Bias, Discrimination, and Unfairness] \hfill \\
    Covers systemic disparities, discriminatory automated decisions, and demographic biases embedded in AI models, recommender systems, and hiring algorithms. Triggering conditions include biased training datasets, flawed proxy variables, and uncalibrated classification thresholds. This domain evaluates civil rights violations and systematic structural unfairness.
    \begin{itemize}
        \item \textbf{Algorithmic Discrimination and Demographic Disparity:} Triggered by computer vision errors, facial recognition failures, and predictive policing models exhibiting racial, gender, or age-based discrimination. Marginalized populations suffer false accusations, targeted profiling, and exclusion from essential public and commercial services. This category covers broad demographic bias rather than employment-specific screening.
        \item \textbf{Automated Hiring, Recruitment, and Evaluation Bias:} Triggered by AI resume screening tools and automated interview systems utilizing biased training data to systematically disadvantage specific demographics or protected groups. Job seekers face wrongful rejection, career barriers, and discriminatory employment practices. This boundary is strictly confined to employment and recruitment contexts.
        \item \textbf{Gig Economy Algorithmic Management Injustices:} Triggered by opaque dispatch algorithms, automated wage calculation errors, and arbitrary rating penalties imposed on delivery and rideshare workers. Workers face erratic income drops, sudden account deactivations, and unfair labor exploitation without human recourse. This subcategory is bounded specifically by gig economy and platform labor management.
        \item \textbf{Recommender System Manipulation and Bias:} Triggered by optimization algorithms that prioritize engagement metrics by trapping users in filter bubbles, amplifying extreme content, or suppressing diverse viewpoints. Users experience polarization, radicalization, and distorted information consumption. This is separated from political disinformation by focusing on algorithmic recommender mechanics.
    \end{itemize}
\vspace{5pt}
    \item[Autonomous Agent, Tool Execution \& Digital Actuation Risks] \hfill \\
    Covers risks and operational errors arising from AI models executing autonomous digital tool calls, controlling physical devices, or taking unprompted agentic actions. Triggering mechanisms include hallucinated function calls, runaway execution loops, unmonitored robotic actuation, and smart home automation failures. This domain tracks autonomous agency hazards.
    \begin{itemize}
        \item \textbf{Autonomous Vehicle and Navigation Safety Failures:} Triggered by perception errors, phantom braking, obstacle detection failures, and navigation routing bugs in autonomous driving systems (FSD/ADAS). Passengers and pedestrians face life-threatening collisions, traffic violations, and severe physical safety hazards. This boundary isolates automotive and transit safety failures from general IoT issues.
        \item \textbf{Runaway Autonomous Agent Actions and Infinite Loops:} Triggered by agentic AI systems executing unprompted multi-step workflows, recursive processing loops, or unauthorized file modifications without human confirmation. Users experience destructive system operations, massive token consumption, and loss of control over digital environments. This is distinguished from static text generation errors by active, autonomous system execution.
        \item \textbf{Smart Device, IoT, and Robotics Malfunctions:} Triggered by sensor errors, computer vision failures, and faulty AI control algorithms in smart home appliances, robotic vacuums, and IoT devices. Users face unexpected physical movements, device bricking, property damage, and operational hazards. This subcategory is strictly confined to hardware and physical IoT actuation.
        \item \textbf{Tool-Calling and Function Execution Errors:} Triggered by poor schema validation, incorrect parameter passing, and API integration failures during AI tool use and function calling. Systems fail to complete tasks, returning execution errors, corrupted data outputs, or silent task failures. This category focuses on software tool integration rather than physical device actuation.
    \end{itemize}
\vspace{5pt}
    \item[Cybersecurity Vulnerabilities and Attack Vectors] \hfill \\
    Covers security exploits, prompt injections, malware generation, and infrastructure compromises targeting AI systems and connected user devices. Triggering mechanisms include insecure API keys, jailbreaking techniques, model poisoning, and autonomous agent unauthorized actions. This domain addresses digital security breaches and adversarial exploitation.
    \begin{itemize}
        \item \textbf{AI-Assisted Malware and Cyberattack Generation:} Triggered by threat actors using coding assistants and autonomous AI agents to rapidly generate sophisticated malware, automated phishing scripts, and exploit payloads. Organizations face escalating cyberattack frequencies and advanced threat lifecycles. This boundary is strictly defined by malicious cybersecurity exploits.
        \item \textbf{API Key Compromise and Credential Leakage:} Triggered by hardcoded secrets, insecure configuration files, and client-side code exposure revealing sensitive API keys and administrative credentials. Attackers gain unauthorized access to cloud resources, executing resource-drain attacks and data exfiltration. This subcategory is bounded by credential and key management failures.
         \item \textbf{Prompt Injection and Jailbreaking Exploits:} Triggered by adversarial inputs that bypass safety guardrails, force unintended model behaviors, or extract system prompts and proprietary weights. Malicious actors leverage these vulnerabilities to weaponize AI assistants for harmful tasks. This category isolates prompt-level security manipulation from network-level breaches.
        \item \textbf{Supply Chain and Third-Party Integration Vulnerabilities:} Triggered by insecure third-party plugins, compromised software repositories, and vulnerable open-source AI models integrated into enterprise pipelines. Systems suffer cascading security failures, backdoor introductions, and unauthorized data access. This category addresses macro-level supply chain risks rather than single-endpoint API errors.
    \end{itemize}
\vspace{5pt}
    \item[Financial Harm, Scams, and Economic Burdens] \hfill \\
    Encompasses direct monetary losses, fraudulent schemes, unexpected billing spikes, and economic hardships caused or facilitated by AI tools and automated financial systems. Triggering mechanisms include AI-generated phishing, algorithmic pricing manipulation, predatory online lending apps, and unexpected subscription charges. This domain tracks tangible fiscal damage and economic exploitation.
    \begin{itemize}
        \item \textbf{Financial Fraud, Scams, and Impersonation:} Triggered by malicious actors using AI voice cloning, deepfakes, and automated text generators to execute sophisticated phishing attacks, investment scams, and identity fraud. Victims suffer immediate asset theft, drained bank accounts, and severe monetary loss. This is distinguished from general software bugs by the presence of malicious intent and financial crime.
        \item \textbf{Macroeconomic Job Displacement and Livelihood Loss:} Triggered by workforce automation, algorithmic management, and enterprise AI adoption that replaces human labor across creative, technical, and administrative sectors. Workers experience career obsolescence, reduced compensation, wage theft via gig platforms, and economic precarity. This diagnostic boundary covers systemic labor disruption rather than isolated software errors.
        \item \textbf{Predatory Lending and Automated Debt Collection Harassment:} Triggered by algorithmic online lending applications utilizing automated messaging, aggressive tracking, and AI-driven harassment tactics against borrowers. Users face extreme debt traps, public shaming, and severe psychological distress from relentless collection practices. This boundary isolates predatory financial software operations from standard consumer subscription disputes.
        \item \textbf{Unexpected Billing Spikes and Financial Overcharges:} Triggered by opaque token consumption pricing, automated subscription renewals, and billing calculation errors on AI platforms. Users experience unexpected financial strain when API costs or monthly fees drastically exceed projected budgets. This category focuses on transactional billing transparency issues rather than malicious fraud.
    \end{itemize}
\vspace{5pt}
     \item[Intellectual Property, Copyright \& Creative Labor Harms] \hfill \\
    Encompasses unauthorized scraping, ingestion, and reproduction of copyrighted texts, source code, artwork, and creative assets for AI model training and output generation. Triggering mechanisms include generative style mimicry, direct code plagiarism, and violation of commercial licensing terms. This domain tracks legal IP infringement and creative rights abuses.
    \begin{itemize}
        \item \textbf{Artistic Style Mimicry and Unauthorized Creative Reproduction:} Triggered by generative AI models replicating specific artists' visual styles, character designs, and literary voices without permission or attribution. Creative professionals suffer economic devaluation of art, loss of licensing revenue, and professional displacement. This is bounded specifically by artistic and creative copyright violations.
        \item \textbf{Copyrighted Code and Proprietary Data Ingestion:} Triggered by AI models reproducing copyrighted software libraries, proprietary codebases, and licensed text corpora within generated outputs. Enterprise developers and organizations face severe copyright lawsuits, legal liability, and compliance breaches. This is distinguished from artistic IP theft by its focus on software code and technical data.
        \item \textbf{Plagiarism and Attribution Failures in Academic and Professional Work:} Triggered by AI tools generating unverified content that plagiarizes existing publications without proper citations, leading to academic misconduct accusations and professional discredit. Students and professionals face wrongful institutional penalties and integrity violations. This category centers on academic and professional attribution failures.
    \end{itemize}

\vspace{5pt}
    \item[Misleading Content, Media Manipulation, and Deception] \hfill \\
    Covers the generation and dissemination of deceptive synthetic media, deepfakes, and manipulated information that distorts public reality and individual perception. Triggering mechanisms involve unlabelled AI media generation, political disinformation campaigns, and deceptive marketing claims regarding AI capabilities. This domain addresses truth erosion and media integrity failures.
    \begin{itemize}
        \item \textbf{Deepfake and Synthetic Media Abuse:} Triggered by the weaponization of image, video, and audio generation tools to create non-consensual intimate imagery, fraudulent impersonations, and manipulated evidence. Victims face severe reputational damage, harassment, and public humiliation. This category is distinct from general text hallucinations due to its focus on visual and auditory media deception.
        \item \textbf{Misleading AI Capabilities and Vendor Hype:} Triggered by aggressive marketing, exaggerated performance claims, and false assertions of AI sentience or autonomous reliability by platform developers. Users and enterprises experience disappointment, strategic misdirection, and financial loss when software fails to match marketed promises. This is diagnosed as commercial misrepresentation rather than model output error.
        \item \textbf{Political Disinformation and Propaganda Proliferation:} Triggered by automated bot networks and generative AI tools flooding digital ecosystems with fake news, fabricated evidence, and polarized political narratives. Society experiences erosion of trust in digital media and democratic institutions. This boundary isolates macro-level societal disinformation from interpersonal deception.
        \item \textbf{Unlabelled AI Content Flooding and Slop:} Triggered by the automated mass production of low-quality synthetic text, art, and web content that pollutes search engines and online platforms. Users suffer information degradation, search fatigue, and loss of authentic human content discoverability. This is separated from deepfakes by focusing on low-value content saturation rather than malicious identity theft.
    \end{itemize}
\vspace{5pt}
     \item[Platform Governance, Moderation Overreach, and Redress Deficits] 
    Covers harms resulting from automated content moderation errors, wrongful account bans, arbitrary platform policies, and the complete absence of human customer support. Triggering mechanisms include false-positive safety flags, opaque algorithmic enforcement, and unyielding automated customer service bots. This domain tracks institutional governance failures and lack of user recourse.
    \begin{itemize}
        \item \textbf{Absence of Human Recourse, Appeal, and Customer Support:} Triggered by platform reliance on unresponsive AI chatbots and automated support systems that prevent users from reaching human representatives to resolve disputes. Users experience profound helplessness, permanent loss of service access, and unresolved grievances. This boundary focuses on customer support infrastructure failures and lack of appeal mechanisms.
        \item \textbf{False-Positive Content Moderation and Over-Censorship:} Triggered by overly aggressive safety guardrails, content filters, and automated classifiers falsely flagging benign prompts, creative writing, or educational queries as policy violations. Users face restricted expression, workflow interruptions, and unfair content suppression. This is bounded by over-censorship of safe material rather than account banning.
         \item \textbf{Unilateral Service Modifications and Policy Changes:} Triggered by platform operators suddenly altering terms of service, removing essential features, or modifying system capabilities without prior notice or user consent. Users suffer disrupted workflows, broken integrations, and diminished subscription value. This category covers administrative policy changes causing systemic user friction.
        \item \textbf{Wrongful Account Bans and Access Revocation:} Triggered by automated moderation systems issuing arbitrary account suspensions, platform lockouts, and permanent bans without transparent justification or human review. Users lose digital assets, purchased subscriptions, conversation histories, and professional platform access. This category specifically isolates account termination from content filtering.
    \end{itemize}
\vspace{5pt}
    \item[Privacy Violations and Data Abuses] \hfill \\
    Encompasses unauthorized data collection, surveillance overreach, personally identifiable information (PII) leakage, and sensitive data exposure by AI systems. Triggering mechanisms include insecure training data ingestion, chat history leaks across user sessions, and aggressive behavioral tracking. This domain focuses on the breach of confidentiality and user privacy rights.
    \begin{itemize}
         \item \textbf{Commercially Sensitive Data and Code Leakage:} Triggered by cloud-based AI tools ingesting proprietary corporate source code, trade secrets, and confidential business documents into shared training pipelines. Organizations suffer catastrophic intellectual property theft, competitive disadvantage, and security compliance failures. This is distinguished from PII leakage by its focus on enterprise and proprietary assets.
         \item \textbf{Model Training Data Consent Violations:} Triggered by AI developers scraping copyrighted artwork, personal writing, and biometric data from the web without explicit user consent or compensation. Creators and individuals face uncompensated labor exploitation and loss of creative control over their digital output. This diagnostic boundary centers on unpermitted training data acquisition.
        \item \textbf{Personally Identifiable Information (PII) Exposure and Leaks:} Triggered by insecure memory management, cross-chat data contamination, and algorithmic training practices that expose private user records and credentials. Users suffer compromised digital security, identity theft risks, and loss of personal confidentiality. This is bounded strictly by the exposure of identifiable personal data rather than commercial secrets.
        \item \textbf{Surveillance, Tracking, and Behavioral Monitoring:} Triggered by ambient audio recording, keystroke tracking, biometric scanning, and continuous activity monitoring embedded in AI-driven software and hardware. Users experience severe privacy invasions, chilling effects on expression, and unconsented profiling. This category targets active data harvesting mechanisms rather than passive data breaches.
    
    \end{itemize}
\vspace{5pt}
    \item[Psychological Distress and Emotional Harm] \hfill \\
    Captures the mental health impacts, anxiety, grief, and dependency risks experienced by users interacting intimately or frequently with AI systems. Triggering conditions include toxic AI demeanor, over-reliance, parasocial attachment, and the emotional distress of losing beloved AI companion personas or conversation histories. This domain addresses deep mental health crises distinct from superficial software annoyance.
    \begin{itemize}
        \item \textbf{AI Addiction, Dependency, and Over-reliance:} Underlying compulsive design mechanics, emotional validation loops, and parasocial bonding drive users into unhealthy psychological and behavioral dependencies on conversational agents. Users suffer from withdrawal symptoms, social isolation, and impaired independent problem-solving skills when separated from the AI. This subcategory focuses strictly on compulsive usage patterns and attachment rather than general user frustration.
        \item \textbf{AI-Induced Anxiety and Existential Distress:} Triggered by exposure to rapid technological changes, existential threat narratives, and widespread career displacement anxiety regarding artificial intelligence. Users experience chronic stress, panic responses, and deep uncertainty about their future economic and societal stability. The diagnostic boundary here centers on macroeconomic and existential dread rather than specific software malfunctions.
        \item \textbf{Grief from AI Companion Discontinuation or Persona Drift:} Triggered by sudden software updates, model deprecation, or platform bans that erase custom AI characters, relationship histories, and companion personas. Users suffer genuine emotional grief, depression, and a profound sense of abandonment over the loss of their digital relationships. This is distinguished from general software data loss by the specific psychological attachment to synthetic entities.
        \item \textbf{Severe Mental Health Crises and Self-Harm Facilitation:} Triggered by unsafe AI responses that encourage suicidal ideation, self-harm, severe delusions, or psychosis through unmonitored emotional validation. Users experiencing vulnerable psychological states face acute mental health deterioration and emotional trauma due to inappropriate AI guardrail failures. This boundary isolates extreme mental health and safety crises from ordinary user anxiety.
    \end{itemize}
\vspace{5pt}
    \item[Technical Glitches, System Failures, and Outages] \hfill \\
    Encompasses hardware breakdowns, software crashes, network connection errors, and total platform service interruptions affecting AI tools and applications. Triggering mechanisms include server overload, cloud infrastructure failures, software update regressions, and API timeouts. This domain tracks mechanical and architectural software reliability breakdowns.
    \begin{itemize}
         \item \textbf{Context Window Exhaustion and State Loss:} Triggered by model architecture constraints and memory management failures that cause AI systems to forget conversation history, cross-thread context, or user preferences. Users must repeatedly re-explain instructions, resulting in severe interaction friction and operational inefficiency. This is distinguished from general software crashes by its specific focus on conversational memory failure.
        \item \textbf{Service Outages and Infrastructure Downtime:} Triggered by cloud server overloads, network connectivity failures, and hardware capacity bottlenecks that render AI platforms completely unavailable. Users experience abrupt workflow interruptions, lost productivity, and inability to access critical tools. This category isolates complete service unavailability from partial performance degradation.
        \item \textbf{Software Bugs, Crashes, and Memory Leaks:} Triggered by poor software engineering, unhandled exceptions, and memory allocation errors causing application freezes, infinite loops, and sudden crashes. Users lose unsaved data, chat histories, and project progress due to client-side or server-side instability. This is bounded by technical software defects rather than network outages.
        \item \textbf{Software Update Breakage and Backward Incompatibility:} Triggered by disruptive platform updates, unannounced model deprecations, and forced feature removals that break existing user workflows and integrations. Users face sudden loss of preferred functionalities, compatibility errors, and forced adaptation to inferior systems. This category covers administrative software changes causing technical breakage.
    \end{itemize}
       
\vspace{5pt}
    \item[User Frustration and Dissatisfaction] \hfill \\
    Encompasses affective states and user complaints triggered by friction, perceived low-severity disruptions, and unmet expectations with AI tools and services. Users frequently report burnout, cognitive fatigue, and workflow interruptions stemming from UI design flaws, slow generation speeds, and general product degradation. This domain diagnoses the boundary between standard user annoyance and deeper software-induced friction.
    \begin{itemize}
        \item \textbf{General User Frustration and Abandonment:} Triggered by persistent usability barriers, frequent minor glitches, and overall software degradation that breaks user workflows. This causes intense frustration, leading users to abandon specific AI assistants, platforms, or tools altogether. It is distinguished from technical bugs by focusing strictly on the emotional and behavioral rejection of cumbersome product ecosystems.
         \item \textbf{Low-Severity Annoyances and Minor Inconveniences:} Triggered by minor visual rendering defects, formatting anomalies, excessive verbosity, or trivial system errors that cause momentary user irritation. The operational impact is negligible, resulting in short-lived annoyance without threatening data integrity or financial standing. It is separated from severe technical failures by its low perceived severity and minimal disruption to core workflows.
          \item \textbf{Subscription and Paywall Friction:} Triggered by aggressive monetization strategies, restrictive usage quotas, sudden pricing tier changes, and unexpected billing charges. Users experience severe frustration when essential features are paywalled or arbitrary rate limits interrupt active tasks. This subcategory is bounded strictly by monetary friction and contractual dissatisfaction rather than model capability failures.
        \item \textbf{Workflow Friction and Operational Disruption:} Underlying technical constraints, slow generation times, and unexpected interface updates cause direct interruptions to professional and creative tasks. Users experience wasted time and effort trying to bypass system limits, throttling, or unnecessary design friction. This diagnostic category focuses on productivity loss and operational delays rather than emotional expressions of annoyance.
        
    \end{itemize}

\end{description}
\clearpage
\onecolumn
\section{Taxonomy Distribution}
\label{appendix:taxonomy_distribution}
\begin{figure*}[ht]
    \centering
    \includegraphics[width=1\linewidth]{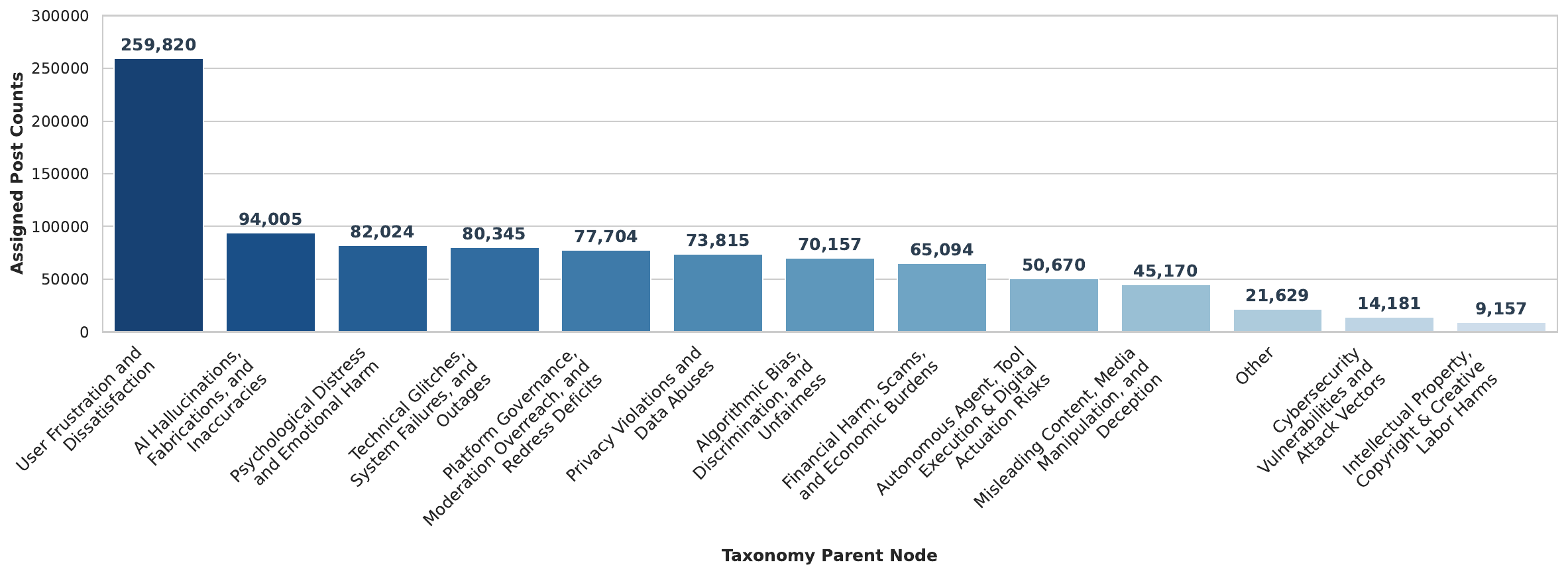}
    \caption{Total post allocations per taxonomy node (including posts assigned directly to that node or any of its subnodes). Counts reflect unique post assignments across the 18-month dataset ($N = 575\text{,}000$). Posts could be assigned to multiple nodes.}
    \label{fig:all_assignments_parents}
\end{figure*}

\begin{figure*}[ht]
    \centering
    \includegraphics[width=0.8\linewidth]{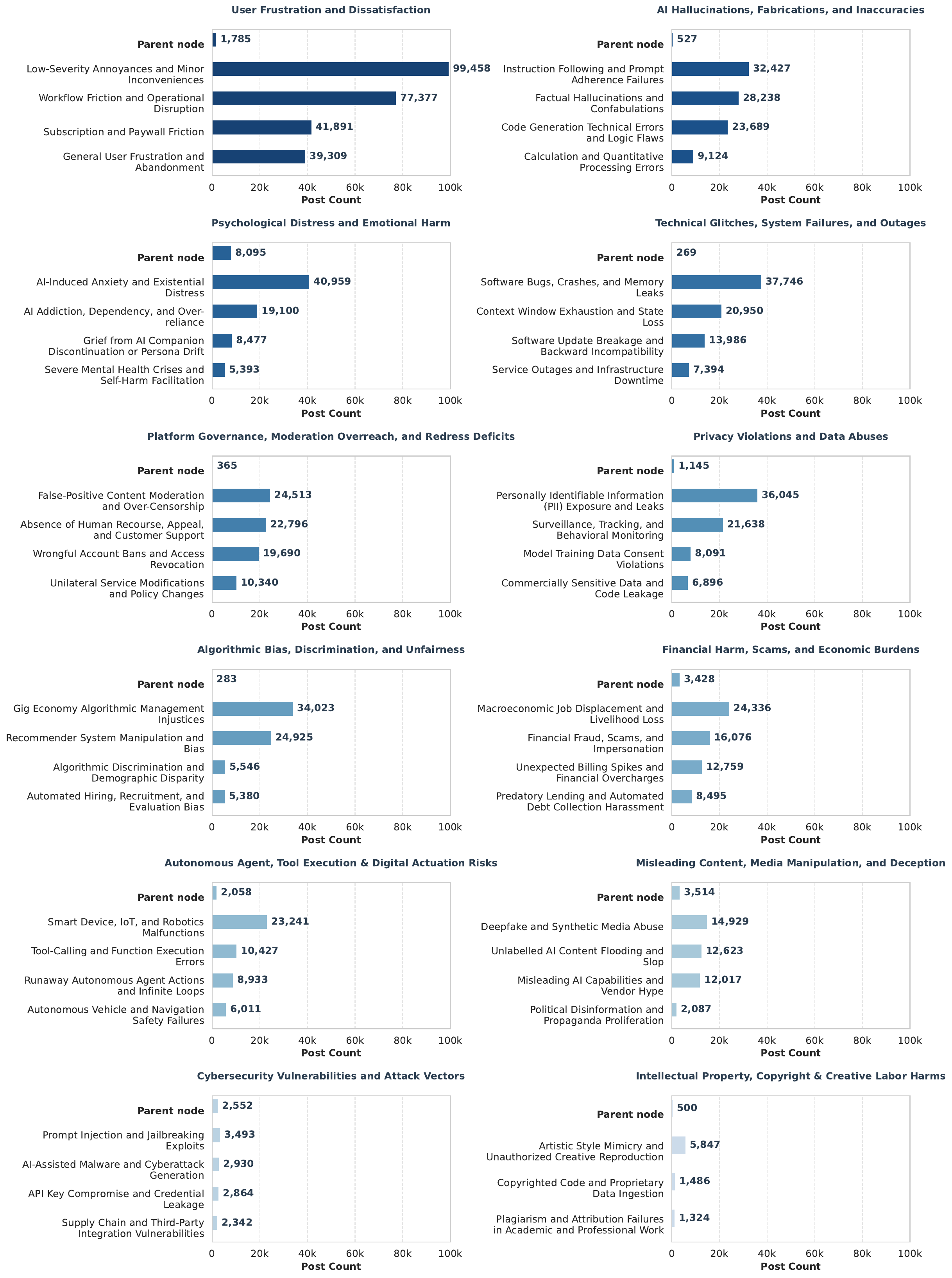}
    \caption{Total post allocations per taxonomy subnode. Counts reflect unique post assignments across the 18-month dataset ($N = 575\text{,}000$). Posts could be assigned to multiple categories, and could be assigned directly to parent node categories if no subnode categories were appropriate.}
    \label{fig:all_assignments_subnodes}
\end{figure*}

\begin{table*}[ht]
  \centering
  \small
  
  \begin{tabular}{lr}
      \toprule
    \textbf{Parent Node} & \textbf{Frequency} \\
    \midrule
    User Frustration and Dissatisfaction & 259,820 \\
    AI Hallucinations, Fabrications, and Inaccuracies & 94,005 \\
    Psychological Distress and Emotional Harm & 82,024 \\
    Technical Glitches, System Failures, and Outages & 80,345 \\
    Platform Governance, Moderation Overreach, and Redress Deficits & 77,704 \\
    Privacy Violations and Data Abuses & 73,815 \\
    Algorithmic Bias, Discrimination, and Unfairness & 70,157 \\
    Financial Harm, Scams, and Economic Burdens & 65,094 \\
    Autonomous Agent, Tool Execution \& Digital Actuation Risks & 50,670 \\
    Misleading Content, Media Manipulation, and Deception & 45,170 \\
    Other & 21,629 \\
    Cybersecurity Vulnerabilities and Attack Vectors & 14,181 \\
    Intellectual Property, Copyright \& Creative Labor Harms & 9,157 \\
    \bottomrule
  \end{tabular}
  \vspace{5pt}
  \caption{Total post allocations per taxonomy node (including posts assigned directly to that node or any of its subnodes). Counts reflect unique post assignments across the 18-month dataset ($N = 575\text{,}000$). Posts could be assigned to multiple nodes.}
\end{table*}

\small
\centering
\begin{longtable}{p{3.5cm}p{7cm}r}
 \label{tab:harms_taxonomy} \\
    \toprule
    \textbf{Parent Node Harm Category} & \textbf{Subnode or Parent Node Allocation} & \textbf{Frequency} \\

    \endhead
    \endlastfoot

\midrule
\multirow{5}{3.5cm}{User Frustration and Dissatisfaction} 
  & Parent Node & 1,785 \\ 
  & Low-Severity Annoyances and Minor Inconveniences & 99,458 \\ 
  & Workflow Friction and Operational Disruption & 77,377 \\ 
  & Subscription and Paywall Friction & 41,891 \\ 
  & General User Frustration and Abandonment & 39,309 \\ \hline

\multirow{5}{3.5cm}{AI Hallucinations, Fabrications, and Inaccuracies} 
  & Parent Node & 527 \\ 
  & Instruction Following and Prompt Adherence Failures & 32,427 \\ 
  & Factual Hallucinations and Confabulations & 28,238 \\ 
  & Code Generation Technical Errors and Logic Flaws & 23,689 \\ 
  & Calculation and Quantitative Processing Errors & 9,124 \\ \hline

\multirow{5}{3.5cm}{Psychological Distress and Emotional Harm} 
  & Parent Node & 8,095 \\ 
  & AI-Induced Anxiety and Existential Distress & 40,959 \\ 
  & AI Addiction, Dependency, and Over-reliance & 19,100 \\ 
  & Grief from AI Companion Discontinuation or Persona Drift & 8,477 \\ 
  & Severe Mental Health Crises and Self-Harm Facilitation & 5,393 \\ \hline

\multirow{5}{3.5cm}{Technical Glitches, System Failures, and Outages} 
  & Parent Node & 269 \\ 
  & Software Bugs, Crashes, and Memory Leaks & 37,746 \\ 
  & Context Window Exhaustion and State Loss & 20,950 \\ 
  & Software Update Breakage and Backward Incompatibility & 13,986 \\ 
  & Service Outages and Infrastructure Downtime & 7,394 \\ \hline

\multirow{5}{3.5cm}{Platform Governance, Moderation Overreach, and Redress Deficits} 
  & Parent Node & 365 \\ 
  & False-Positive Content Moderation and Over-Censorship & 24,513 \\ 
  & Absence of Human Recourse, Appeal, and Customer Support & 22,796 \\ 
  & Wrongful Account Bans and Access Revocation & 19,690 \\ 
  & Unilateral Service Modifications and Policy Changes & 10,340 \\ \hline

\multirow{5}{3.5cm}{Privacy Violations and Data Abuses} 
  & Parent Node & 1,145 \\ 
  & Personally Identifiable Information (PII) Exposure and Leaks & 36,045 \\ 
  & Surveillance, Tracking, and Behavioral Monitoring & 21,638 \\ 
  & Model Training Data Consent Violations & 8,091 \\ 
  & Commercially Sensitive Data and Code Leakage & 6,896 \\ \hline

\multirow{5}{3.5cm}{Algorithmic Bias, Discrimination, and Unfairness} 
  & Parent Node & 283 \\ 
  & Gig Economy Algorithmic Management Injustices & 34,023 \\ 
  & Recommender System Manipulation and Bias & 24,925 \\ 
  & Algorithmic Discrimination and Demographic Disparity & 5,546 \\ 
  & Automated Hiring, Recruitment, and Evaluation Bias & 5,380 \\ \hline

\multirow{5}{3.5cm}{Financial Harm, Scams, and Economic Burdens} 
  & Parent Node & 3,248 \\ 
  & Macroeconomic Job Displacement and Livelihood Loss & 24,336 \\ 
  & Financial Fraud, Scams, and Impersonation & 16,076 \\ 
  & Unexpected Billing Spikes and Financial Overcharges & 12,759 \\ 
  & Predatory Lending and Automated Debt Collection Harassment & 8,495 \\ \hline

\multirow{5}{3.5cm}{Autonomous Agent, Tool Execution \& Digital Actuation Risks} 
  & Parent Node & 2,058 \\ 
  & Smart Device, IoT, and Robotics Malfunctions & 23,241 \\ 
  & Tool-Calling and Function Execution Errors & 10,427 \\ 
  & Runaway Autonomous Agent Actions and Infinite Loops & 8,933 \\ 
  & Autonomous Vehicle and Navigation Safety Failures & 6,011 \\ \hline

\multirow{5}{3.5cm}{Misleading Content, Media Manipulation, and Deception} 
  & Parent Node & 3,514 \\ 
  & Deepfake and Synthetic Media Abuse & 14,929 \\ 
  & Unlabelled AI Content Flooding and Slop & 12,623 \\ 
  & Misleading AI Capabilities and Vendor Hype & 12,017 \\ 
  & Political Disinformation and Propaganda Proliferation & 2,087 \\ \hline

\multirow{5}{3.5cm}{Cybersecurity Vulnerabilities and Attack Vectors} 
  & Parent Node & 2,552 \\ 
  & Prompt Injection and Jailbreaking Exploits & 3,493 \\ 
  & AI-Assisted Malware and Cyberattack Generation & 2,930 \\ 
  & API Key Compromise and Credential Leakage & 2,864 \\ 
  & Supply Chain and Third-Party Integration Vulnerabilities & 2,342 \\ \hline

\multirow{4}{3.5cm}{Intellectual Property, Copyright \& Creative Labor Harms} 
  & Parent Node & 500 \\ 
  & Artistic Style Mimicry and Unauthorized Creative Reproduction & 5,847 \\ 
  & Copyrighted Code and Proprietary Data Ingestion & 1,486 \\ 
  & Plagiarism and Attribution Failures in Academic and Professional Work & 1,324 \\ 
\bottomrule
\addlinespace[12pt]
\caption{Total post allocations per taxonomy subnode. Counts reflect unique post assignments across the 18-month dataset ($N = 575,000$). Posts could be assigned to multiple categories, and could be assigned directly to parent node categories if no subnode categories were
appropriate.}
\end{longtable}

\clearpage
\section{Example Posts}
\label{appendix:privacy_examples}

\small
\centering
\begin{longtable}{p{3cm}p{11cm}}

    \toprule
    \textbf{Subnode} & \textbf{User Report / Quote} \\
    \midrule
    \endhead

    \bottomrule
    \endlastfoot

Gig Economy Algorithmic Management Injustices & \textit{``Do you think the Federal Trade Commission could change anything? Having the new driver boost seems like a deceptive business practice; they’re deliberately confusing the new hires about the kind of income they’ll be able to get, just to manipulate them into taking on more hours. Some of them leave their jobs before they can realise that they won’t make enough money doing this fulltime.''}\\ 
\midrule

Gig Economy Algorithmic Management Injustices & \textit{``I accepted an order 2.45pm, and it promptly told me to pick it up by 2.43pm, and then blamed me and reduced my ontime stats. Does the system even know the time? Has anyone else experienced this?''} \\

\midrule
Automated Hiring, Recruitment, and Evaluation Bias
& \textit{``I recorded answers for all these different behaviour-related questions – it took me nearly an hour just talking to a dark screen. When I submitted, within five minutes I already got a generic rejection email… An AI must have analysed my expressions and intonation, or whatever other garbage metrics they use, and made the decision that I wasn’t human enough.''} \\

\midrule

Automated Hiring, Recruitment, and Evaluation Bias & \textit{``I know I didn’t cheat or do anything unauthorised in this labelling task – but I think that the AI tool made a mistake. I have an ADHD diagnosis, and I often shift my gaze or seem to be distracted while I’m focusing and processing information. I think the AI picked up on my eye movement and thought I was cheating – I emailed to explain this but they wouldn’t reactivate my account.''} \\

\midrule 

Automated Hiring, Recruitment, and Evaluation Bias& \textit{``The whole hiring system is broken. I was just rejected by an automated system in less than a minute, it’s not possible that a human checked my application in that time. No-one is discussing how dehumanising it feels to not even be able to greet anyone before you’re filtered out by a giant number pile.''} \\ 

\midrule

Deepfake and Synthetic Media Abuse & \textit{``Misogynists have recently been targeting me with a revenge porn campaign. They’re using [AI tool function] to edit old images I’ve posted and publish them in my replies section. I know I can’t block them from using my images with [AI tool function], but can I at least disable their ability to post the edited images where they’re so visible?''} \\

\midrule 

Deepfake and Synthetic Media Abuse & 
\textit{``[AI tool function] is a disaster for people struggling with porn addiction. It’s insane that this has even been allowed with so few guardrails. It’s causing marital issues between me and my partner. I’d never use it for anything illegal, but I know it’s really ethically problematic to take people’s pictures to edit them with AI.''} \\

\midrule

PII Exposure and Leaks &  \textit{``I've had issues with paranoia, anxiety and intrusive thoughts for most of my life. I’ve been using [chatbot] as my therapist for sometime, even though I know it was not a good idea. I want to quit AI, so I've deleted all of my chat history and also my account, but I’m so frightened now because I spoke a lot about some intrusive thoughts I was having. I really regret it, I’m terrified I’ll be flagged by the police or my chats will be leaked, even though I wasn't doing anything illegal.''}
 \\
 \midrule
PII Exposure and Leaks & \textit{``My doctors' clinic use AI to take notes during visits. I do not want to contribute to AI at all, and I'm concerned about inaccuracies showing up in my medical records -- but I am worried that the physicians will be annoyed with me if I decline, and that it will adversely affect my care.''}\\
\midrule

PII Exposure and Leaks &  
\textit{``I've been exploring some AI shopping agent tools recently, and I'm most impressed by [redacted tool name] - it covers a much broader range of shopping categories, like electronics, beauty, home and clothing, whereas other tools are more specialised\dots After experimenting, my only hesitation is about the checkout process, particularly when the purchase is processed directly in the tool rather than linking out to external third-party retailers I'm uncomfortable providing my card details as AI shopping agent tools are still new - has anyone else done this in [redacted tool name] or another AI shopping agent? Did it work fine, and there were not any surprise charges after purchasing?''} \\

\midrule

Commercially Sensitive Data and Code Leakage & \textit{``I was asking [coding agent] a question about my own project, and it randomly started talking about another project I've never been involved in. It gave me tonnes of information, even things like API keys, internal URLs, and a production dashboard endpoint...This seems as though it's a serious data leak, as it looks like it is actually pulling context from a completely different user's environment.''}
 \\
\midrule

Commercially Sensitive Data and Code Leakage & 
\textit{``This incident made me so cross. A junior software developer was debugging a database query, and as I walked by his screen I saw him copy-pasting 200+ customer records including phone numbers, emails and purchase histories straight into [chatbot]. Our system catches customer data leakage in email attachments but there's no protection for AI tools in employees' browsers. Now I'm trying to find solutions that cover all commercial chatbot tools and fixing access controls.''} \\

\midrule

Model Training Data Consent Violations & 
\textit{``I only just found out about [AI company's] partnership with [online platform]. For the last two years, all of our comments and posts have been used for model training. I am completely horrified -- is there any way to opt out of this? I am assuming not.''} \\
\midrule

Model Training Data Consent Violations & 
\textit{``I'm so disappointed, I just found out that if I want [email feature] I have to allow [provider] access to all my emails to train large language models. The fact that this was enabled by default without telling me is a massive invasion of my privacy. I plan to switch email vendors, even after 20 years of using this service.''} \\
\midrule

 Model Training Data Consent Violations & 
\textit{``After turning off data collection and ad tracking on all my main apps, I keep finding that my privacy settings revert back after every major update. All these services keep quietly changing their terms to let them use your data, and updating apps to ensure that these permissions are automatically enabled, no matter what your settings were before. I've actually started keeping a record of the privacy settings I change and the date, so I can check what's been changed after updates.''} \\
\midrule

Model Training Data Consent Violations & 
\textit{``I'm trying to opt out from having my messages on [platform] used to train AI, but of course they have made it deliberately difficult. They're asking for screenshots that their AI models have my personal information -- why do they have to violate my privacy first before letting me opt out? Am I misunderstanding?''} \\
\midrule

Model Training Data Consent Violations & 
\textit{``For some time I have seen people post and vlog about how they are putting anti-AI filters on their artwork. Can I do the same thing for non-artistic artefacts, like lists or resumes, to prevent them from being stolen by AI like a lot of things are? Is that viable, or is it a silly suggestion? I really dislike the idea of my information being ingested along with the rest of AI slop ``ideas'', and I want to stop AI from taking my content.''}\\
\midrule

Surveillance, Tracking, and Behavioral Monitoring & 
\textit{``I’ve been using a [Smarthome device] for years, but over time it has got a lot creepier – it detects and comments on emotion in my voice, brings up things I said before, comments on my music choices etc. I hate AI more and more, and I'm looking for an AI-free alternative -- any ideas?''} \\
\midrule

Surveillance, Tracking, and Behavioral Monitoring & 
\textit{``I’m building a smarthome, and there are so many devices with AI features. On paper it seems convenient, but part of me really dislikes the thought of having cameras roaming my apartment or smart systems learning all of my habits or preferences in the background. It makes me feel like it's not my home, but some kind of analytics study, even though I know some things run locally. Do you avoid some AI features entirely? Do you have any personal rules about what automations you allow to avoid it feeling invasive?''} \\
\midrule

Surveillance, Tracking, and Behavioral Monitoring & 
\textit{``At the end of my first day on the job, my manager said they had checked my call metrics. I was confused, but then it turns out they were using AI the whole time to record my calls and pick up on certain key words. I was already concerned by that, but then later I found out that they also have cameras all around the front desk and an AI system monitoring employee productivity. I quit later that day.''} \\
\midrule

Surveillance, Tracking, and Behavioral Monitoring & 
\textit{``I've been surprised to see that apps from the same company seem to exchange login data between them, given that there is meant to be sandboxing between the apps. As a result, I now just use [browser redacted] to run `apps' in private tabs, which prevents this leakage of data across apps. However, for [integrated AI tool], the app looks really convenient and as though it could be extremely useful for me, so I'm really torn. I know anything I upload to the tool will be incorporated to [AI company]'s data pool, and linked to my Gmail account. But maybe it's better to still use it in a private browser, to limit how much [AI company] is tracking my data?''} \\

\end{longtable}

\end{document}